\documentclass[pdflatex,sn-nature]{sn-jnl}% Style for submissions to Nature Portfolio journals
\usepackage{graphicx}%
\usepackage{multirow}%
\usepackage{amsmath,amssymb,amsfonts}%
\usepackage{amsthm}%
\usepackage[title]{appendix}%
\usepackage{color}%
\usepackage{xcolor}%
\usepackage{textcomp}%
\usepackage{manyfoot}%
\usepackage{booktabs}%
\usepackage{algorithm}%
\usepackage{algorithmicx}%
\usepackage{algpseudocode}%
\usepackage{listings}%
\usepackage{float}
\usepackage{siunitx}
\DeclareSIUnit\bar{bar}
\usepackage{pgffor}
\usepackage{pdfpages}
\usepackage{multicol}
\usepackage{svg}

\usepackage{ulem}
\usepackage{comment}

\theoremstyle{thmstyleone}%
\theoremstyle{thmstyletwo}%

\theoremstyle{thmstylethree}%

\makeatletter
\AtBeginDocument{\let\LS@rot\@undefined}
\makeatother

\begin{document}

\title{Nonadiabatic dynamics under slow parametric modulation in nonlinear graphene resonators}

\author*[1]{\fnm{Chris F. D.} \sur{Wattjes}}\email{C.F.D.Wattjes-1@tudelft.nl}
\author[1]{\fnm{Haemin} \sur{Lim}}
\author[2]{\fnm{Pierpaolo} \sur{Belardinelli}}
\author[3]{\fnm{John E.} \sur{Sader}}
\author[1]{\fnm{Peter G.} \sur{Steeneken}}
\author*[1]{\fnm{Farbod} \sur{Alijani}}\email{F.Alijani@tudelft.nl}

\affil[1]{\orgdiv{Department of Precision and Microsystems Engineering}, \orgname{Delft University of Technology}, \orgaddress{\street{Mekelweg 2}, \city{Delft}, \postcode{2628 CD}, \country{The Netherlands}}}
\affil[2]{\orgdiv{Department of Construction, Civil Engineering and Architecture}, \orgname{Polytechnic University of Marche}, \orgaddress{\street{via Brecce Bianche}, \city{Ancona}, \postcode{60131}, \country{Italy}}}
\affil[3]{\orgdiv{Lynn Booth \& Kent Kresa Department of Aerospace \& Department of Applied Physics and Materials Science}, \orgname{California Institute of Technology}, \city{Pasadena}, \postcode{CA 91125}, \country{USA}}

\date{\today}

\abstract{
Resonantly driven nonlinear systems, from superconducting circuits to nanomechanical resonators, are often described in a rotating frame, where amplitude and phase evolve slowly within a fixed landscape of stable and unstable states. How such systems behave when this landscape moves on their intrinsic timescale remains to be understood. Here, we exploit the strong tension sensitivity of a graphene drum’s resonance frequency to reshape the rotating-frame phase space of a bistable mechanical mode in real time, combining an MHz-frequency resonant drive with kHz-frequency tension modulation. When the modulation rate becomes comparable to the mechanical relaxation rate, adiabatic tracking of the fixed points breaks down. Instead, parametric instabilities emerge in the slow dynamics at frequencies set by libration around these points rather than by the mechanical resonance; stronger modulation drives period doubling and chaos. A slow-variable theory predicts where the instabilities occur, and Melnikov analysis captures the onset of chaos. These results show that the rotating frame itself can be parametrically driven, offering a general route to controlling nonadiabatic dynamics in systems ranging from Josephson bifurcation amplifiers to cold atoms.
}

\maketitle

Driven-dissipative nonlinear mesoscopic systems often exhibit dynamics on well-separated timescales. In these systems, the physical motion occurs near the drive frequency, while their amplitude and phase evolve more slowly under the combined effects of dissipation and nonlinearity~\cite{bachtold2022mesoscopic, eichler2023classical}. This separation is conveniently described in a frame rotating with the drive, where the rapid oscillations are averaged out and the slow amplitude–phase dynamics appear as trajectories in phase space, organized by stable and unstable fixed points and the surrounding vector flow \cite{strogatz2024nonlinear}. Nanomechanical resonators provide a particularly accessible realization of this picture, in which slow amplitude–phase dynamics provide a natural framework for describing phenomena including noise squeezing~\cite{Huber2020Spectral}, synchronization~\cite{matheny2014phase,houri2026three}, nonlinear mode coupling~\cite{kecskekler2021tuning,shoshani2021resonant,yan2022energy}, noise-induced transitions~\cite{,Belardinelli2025, boness2026nonequilibrium}, and frequency-comb generation~\cite{czaplewski2018bifurcation,Ochs2022FrequencyComb}. For fixed driving conditions, however, the underlying rotating-frame phase-space structure in these systems typically remains stationary.

A qualitatively different regime arises when this phase-space structure itself evolves in time, most interestingly when the evolution timescale is comparable to the intrinsic relaxation of the system. In micro- and nanomechanical resonators, one route to introducing such an additional timescale is via dual-tone excitation, where two closely spaced drive tones generate a beat frequency that periodically changes the rotating-frame dynamics and can induce transitions between coexisting dynamical states~\cite{Catalini2025,Kumar2026,Eshel2026twotone}. When the beat period is slow compared to the intrinsic relaxation time, the system can continuously adjust and follow its instantaneous state. As the two timescales become comparable, however, adiabatic following can break down. Such timescale competition underlies nonadiabatic dynamics in mesoscopic physics, as demonstrated in quantum dots and superconducting circuits~\cite{ribeiro2013coherent,quintana2013cavity}. In nanomechanical resonators, nonadiabatic behaviour has also been observed when two coupled modes are swept through an avoided crossing on a timescale comparable to their intermodal coupling time~\cite{faust2012nonadiabatic}. More recently, theoretical work has shown that the interplay between slow and fast timescales in multistable nonlinear systems can generate dynamics that are missed by quasistatic or adiabatic approximations~\cite{yanchuk2026singular}. The experimental realization and consequences of such nonadiabatic dynamics in a single nonlinear nanomechanical mode, where the phase-space evolution timescale becomes comparable to the intrinsic relaxation time, remain  unexplored.

In this work, we investigate the nonadiabatic dynamics of a single nonlinear mode of a graphene drum by introducing a second timescale through parametric modulation of an intrinsic resonator property, namely its membrane tension. The strong dependence of graphene’s resonance frequency on membrane tension makes this system particularly well suited for generating a time-evolving nonlinear phase-space. Tension modulation in graphene membranes has previously been exploited to investigate parametric resonance~\cite{Dolleman2018, Kartal2026stability}, the interplay between parametric and direct interactions~\cite{kecskekler2021tuning}, as well as internal resonance~\cite{rathi2026tunablenonlinearlandscapesgraphene}. Here, however, we adopt a distinct approach: rather than using fast tension modulation as the primary drive, we exploit it to slowly reshape the phase-space structure at a frequency much smaller than the resonance frequency while independently maintaining a resonant excitation. To this end, we simultaneously apply a fast MHz direct drive near the fundamental resonance and a slow kHz modulation of the membrane tension. The fast drive establishes a nonlinear bistable response, whereas the slow modulation continuously shifts the resonance frequency, thereby moving the stable and unstable fixed points that define the phase-space in the rotating frame. Therefore, unlike the dual-tone schemes ~\cite{Catalini2025,Kumar2026}, the underlying phase-space is itself directly modulated, not just through external direct forcing. Under the parametric modulation, the rate of phase-space evolution competes directly with the mechanical relaxation rate of the membrane. By varying the frequency and amplitude of the slow modulation, we tune both the rate at which the phase-space evolves  and the degree of distortion of its structure. When the modulation is slow compared to the relaxation rate, the response adiabatically tracks the instantaneous stable states. In the opposite limit, where the modulation is too fast, the response can no longer follow the changes in phase-space and lags behind the moving attractors. Most strikingly, when the modulation rate becomes comparable to the relaxation rate, the system enters a critical regime in which the parametric modulation can activate instabilities in the slow dynamics, giving rise to period-doubling and chaotic motion that dominate the nanomechanical response. Combining experiments and numerical simulations, we discuss the root cause of the observed dynamics and use Melnikov theory to provide a mechanistic picture of the transition to chaos. More broadly, our results show that controlling the rate at which a nonlinear phase-space evolves offers a new degree of freedom for engineering the dynamical response of nanomechanical systems.

%%%%%%%%%%%%%%%%%%%%%%%%%%%%%%%%%%%%%%%%%%%%%%%%%%%%%%%%%%%%%%%%%%%%%%%%%%%%%%%%%
%%%%%%%%%%%%%%%%%%%%%%%%%%%%%%%%%%%%%%%%%%%%%%%%%%%%%%%%%%%%%%%%%%%%%%%%%%%%%%%%%

\begin{figure}
    \centering
    \includegraphics[width=\linewidth]{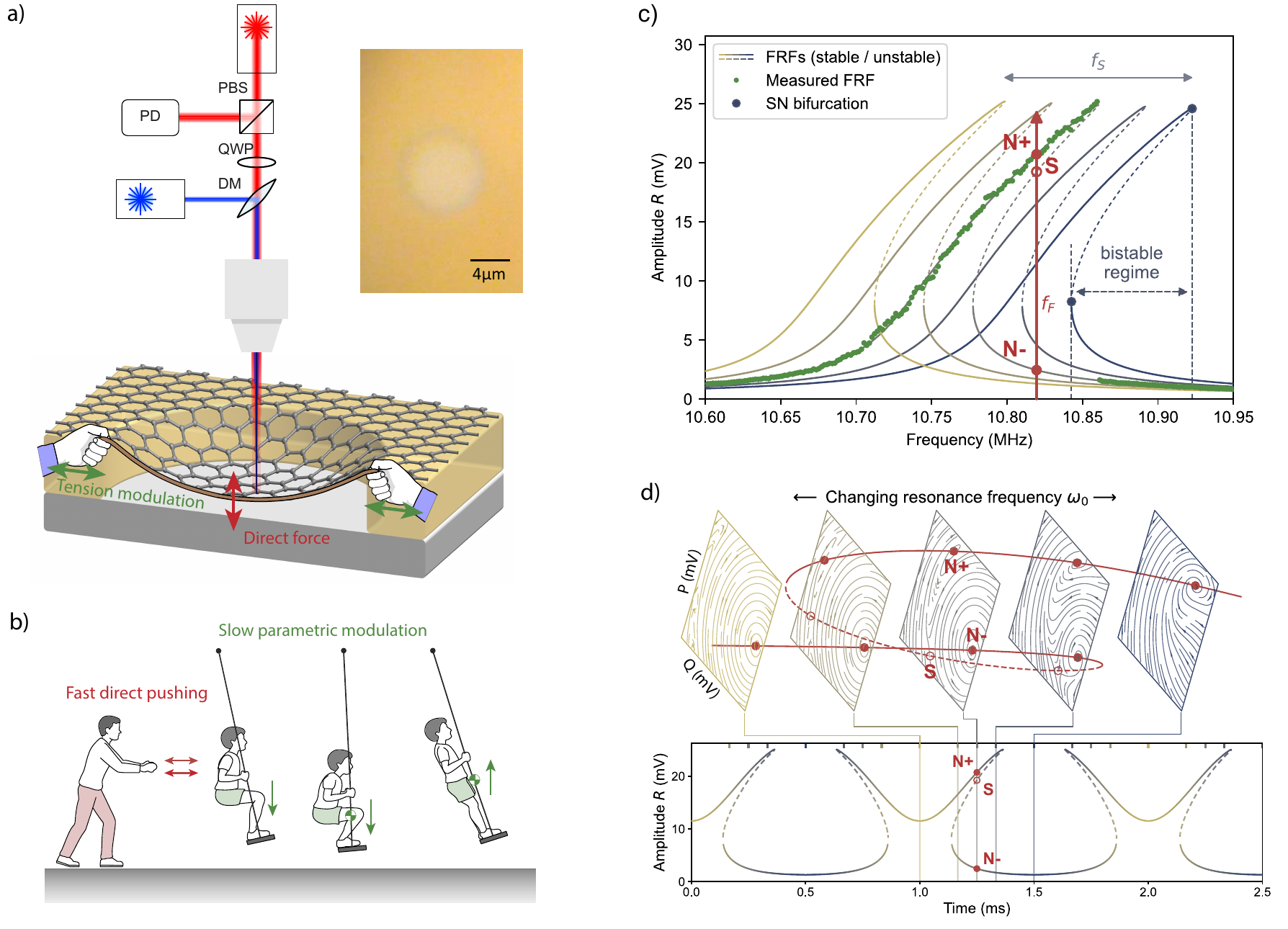}
    \caption{\textbf{Slow parametric--fast direct driving of graphene membranes}. 
\textbf{a}, Fabry--P\'erot interferometry setup (PD: photodiode, PBS: polarizing beam splitter, QWP: quarter-wave plate, DM: dichroic mirror), with a red readout laser and blue drive laser. 
\textbf{b}, Dual-drive concept illustrated by a child-on-a-swing analogy: slow parametric modulation changes the stiffness, while fast direct forcing provides the resonant drive.
\textbf{c}, Nonlinear frequency response (green) and Duffing fit (solid lines) without slow modulation, yielding
$f_0=10.765\pm0.019\,\si{\mega\hertz}$,
$\Gamma=4.90\pm0.81\times10^{4}\,\si{\per\second}$,
$\gamma=2.15\pm0.61\times10^{17}\,\si{\volt^{-2}\second^{-2}}$, and
$F_\mathrm{F}=1.65\pm0.38\times10^{11}\,\si{\volt\second^{-2}}$ at $V_\mathrm{F}=\SI{1.5}{\volt}$.
The slow modulation periodically shifts the resonance at $f_\mathrm{S}$, while $f_\mathrm{F}$ is fixed within the bistable region (red arrow); N$\pm$ and S denote the stable and unstable solutions.
\textbf{d}, Rotating-frame vector fields (top) and fixed-point amplitudes (bottom) as the resonance is varied; colours correspond to panel \textbf{c}.}
    \label{fig:physics-and-explanation}
\end{figure}

\section{Dynamically evolving phase-space}

We consider a nonlinear resonator subjected to two distinct forms of excitation: a fast drive in the spectral neighbourhood of its resonance, and a parametric modulation at much lower frequencies. We implement this scenario experimentally using a graphene drum. The device consists of a chemical vapour deposition (CVD) bilayer graphene membrane suspended over a circular cavity with a diameter of \SI{8}{\micro\meter} and a depth of \SI{420}{\nano\meter} etched into a SiO\textsubscript{2} layer. The membrane displacement is measured interferometrically using a red laser (\SI{632.8}{\nano\meter}), while a blue laser (\SI{488}{\nano\meter}) provides optothermal actuation~\cite{Dolleman2018}. 

A useful analogy to this configuration is a child on a swing. While a parent pushes the swing close to its resonance frequency (direct drive), the child periodically changes the effective stiffness of the system by moving their centre of mass up and down (parametric modulation), as illustrated in Fig.~\ref{fig:physics-and-explanation}a, b. The swing analogy is often used to illustrate parametric resonance: when the child moves at twice the resonance frequency parametric resonance is activated and the oscillation amplitude grows. Here, however, we consider a different situation: the child moves much slower, while their parent pushes the swing close to the natural frequency of the system. Thus, the parent determines the oscillatory motion, whereas the child influences the motion indirectly by slowly shifting the equilibrium positions of this dynamical scenario.

Moving back from the playground to nanomechanics, we modulate the intensity of the blue laser to induce periodic heating and cooling of the membrane. The resulting thermal expansion alters the membrane tension, modifying the effective stiffness of the resonator and thereby providing a parametric drive (see Fig.~\ref{fig:physics-and-explanation}a). 
At the same time, the tension modulation generates a direct force owing to the static out-of-plane deflection of the membrane, which arises commonly from membrane adhesion to the sidewalls of the cavity~\cite{steeneken2021dynamics}. This effect is analogous to a sagged string. For a straight string, changing the tension modifies only the stiffness. However, when the string is statically deflected (as illustrated in Fig.~\ref{fig:physics-and-explanation}a), a tension change also produces a transverse restoring force and thereby directly drives the out-of-plane motion. The static membrane deflection plays the same role in the graphene resonator, converting tension modulation into a direct out-of-plane force.

To realize the combined slow parametric modulation and fast resonant driving, we modulate the blue-laser intensity using $V_{\mathrm{blue}}(t)= V_{\mathrm{S}}\cos(\omega_{\mathrm{S}}t)+V_{\mathrm{F}}\cos(\omega_{\mathrm{F}}t),$
where $V_{\mathrm{S}}$ and $V_{\mathrm{F}}$ are the voltage amplitudes applied to the laser diode at the slow modulation frequency $\omega_{\mathrm{S}}\ll\omega_0$ and the near-resonant frequency $\omega_{\mathrm{F}}\approx\omega_0$, respectively. Before applying the slow modulation, we characterize the resonant response of the drum. At $V_{\mathrm{F}}=1.5\,\mathrm{V}$, the graphene drum exhibits a bistable hardening response (see Fig.~\ref{fig:physics-and-explanation}c), where low- and high-amplitude stable oscillations coexist and are separated by an unstable branch. The fundamental resonance frequency of the drum is found to be  $f_0=10.8\,\mathrm{MHz}$ and its quality factor is $Q=708$. 

The addition of the low-frequency component next introduces a slow modulation of the membrane tension and hence of its resonance frequency. In principle, the two components of $V_{\mathrm{blue}}(t)$ can generate both direct and parametric forcing at the slow and fast frequencies. Introducing the small parameter $\epsilon\ll1$ that sets the order of the static out-of-plane deformation and slow thermal modulation, a scaling analysis (Supplementary Note~2) shows that the fast resonant direct drive and slow tension modulation both enter the equation of motion at $O(\epsilon)$, whereas the slow direct forcing is $O(\epsilon^2)$. The fast tension modulation enters at leading order but is non-resonant because the fast drive is applied near $f_0$, far from the parametric-resonance condition. The resulting dynamics can therefore be captured by a minimal single-mode Duffing-Mathieu model of the form,
\begin{equation}
    \ddot{x} +2\Gamma\dot{x} +\left(\omega_0^2+F_{\mathrm{S}}\cos(\omega_{\mathrm{S}}t)\right)x
    +\gamma x^3= F_{\mathrm{F}}\cos(\omega_{\mathrm{F}}t),
    \label{eq:eom}
\end{equation}
where $x$ denotes the out-of-plane displacement of the membrane, $\Gamma$ is the damping rate, $\omega_0= 2\pi f_0$,  and $\gamma$ is the Duffing-type nonlinear stiffness associated with geometric nonlinearity~\cite{davidovikj2017nonlinear}. The fast forcing amplitude is $F_{\mathrm{F}}$, whereas the slow modulation enters through $F_{\mathrm{S}}\cos(\omega_{\mathrm{S}}t)$ in the effective linear stiffness. Since $F_{\mathrm{S}}\ll\omega_0^2$, this term produces a slow periodic shift of the instantaneous resonance frequency. The parameters $\omega_0$, $\Gamma$, $\gamma$, and $F_{\mathrm{F}}$ are determined from the measured Duffing response (see Supplementary Note~1), while the slow modulation strength is calibrated from the modulation-induced response.

To analyse the resulting nonlinear dynamics, we transform Eq.~\eqref{eq:eom} to the frame rotating at $\omega_{\mathrm{F}}$ and apply the method of averaging~\cite{kryloff1947introduction} (see Supplementary Note~3). Introducing slowly varying in-phase and out-of-phase quadratures, $P(t)$ and $Q(t)$, respectively, and averaging over one fast oscillation period removes the rapidly oscillating terms and reduces the dynamics to a pair of coupled first-order equations,
\begin{equation}
\label{eq:slow-eom}
\left\{
\begin{aligned}
\dot{P} &=
\frac{\hat{\omega}_0^{\,2}-\omega_{\mathrm{F}}^{\,2}}
     {2\omega_{\mathrm{F}}}\,Q
-\Gamma P
+\frac{3}{8}\frac{\gamma}{\omega_{\mathrm{F}}}
Q\left(P^2+Q^2\right),
\\[6pt]
\dot{Q}
&=
-\frac{\hat{\omega}_0^{\,2}-\omega_{\mathrm{F}}^{\,2}}
      {2\omega_{\mathrm{F}}}\,P
-\Gamma Q
-\frac{3}{8}\frac{\gamma}{\omega_{\mathrm{F}}}
P\left(P^2+Q^2\right)
+\frac{F_{\mathrm{F}}}{2\omega_{\mathrm{F}}},
\end{aligned}
\right.
\end{equation}
where the displacement is written as $x(t)=P(t)\cos(\omega_{\mathrm{F}}t)+Q(t)\sin(\omega_{\mathrm{F}}t),$ with vibration amplitude $R=\sqrt{P^2+Q^2}.$
The slow tension modulation appears through the instantaneous resonance frequency, $\hat{\omega}_0(t)=\omega_0+A_{\mathrm{S}}\cos(\omega_{\mathrm{S}}t),$ where $A_{\mathrm{S}}\approx F_{\mathrm{S}}/(2\omega_0)$ is the modulation strength. By linearizing Eq.~\eqref{eq:slow-eom} about the stable slow-flow fixed points $N\pm$ (see Supplementary Note~4), we obtain the corresponding eigenvalues as $\lambda_\pm=-\Gamma\pm i\omega_\mathrm{L}$, where 

\begin{equation}
    \omega_\mathrm{L}=\sqrt{\left(\frac{\omega_0^{\,2}-\omega_{\mathrm{F}}^{\,2}}
      {2\omega_{\mathrm{F}}}+\frac{3}{8}\frac{\gamma}{\omega_{\mathrm{F}}}R_0^2\right)\left(\frac{\omega_0^{\,2}-\omega_{\mathrm{F}}^{\,2}}
      {2\omega_{\mathrm{F}}}+\frac{9}{8}\frac{\gamma}{\omega_{\mathrm{F}}}R_0^2\right)},
      \label{eq:slow-flow-eigenfrequency}
\end{equation}
 is the libration frequency governing small oscillations about the fixed point, and $R_0$ is the amplitude of the stable fixed points. In the absence of nonlinearity, the libration frequency reduces to $\omega_\mathrm{L}\approx|\omega_0-\omega_\mathrm{F}|$. 

The slow harmonic modulation of the tension causes the entire frequency-response curve to move back and forth in the amplitude–frequency space (Fig.~\ref{fig:physics-and-explanation}c). As a result, the vector field evolves continuously in time, rendering the positions and existence of the fixed points, two stable spirals (N$\pm$) and one saddle $\mathrm{S}$, time dependent. For sufficiently strong modulation, the system may periodically cross the saddle-node bifurcations, causing the number of available fixed points to vary on the slow timescale set by $\omega_\mathrm{S}$. Tracing these fixed points throughout a modulation cycle yields the trajectories shown in Fig.~\ref{fig:physics-and-explanation}d, with periodic creation and annihilation of equilibria. 
However, the previous picture of evolving steady-state equilibria does not account for the finite time required for the resonator to reach such states during the time-varying evolution of the phase-space. The actual dynamics are governed by the competition between these two timescales: the rate at which the phase-space topology evolves and the rate at which the resonator relaxes towards the instantaneous fixed points. 

When the modulation and relaxation timescales become comparable, a third characteristic timescale associated with the slow-flow dynamics becomes relevant, which is set by the libration frequency $\omega_\mathrm{L}$. As we show in Supplementary Note~4, slow parametric modulation of the fast dynamics leads to an effective parametric and direct drive in the slow dynamics when linearized around its fixed points. Direct resonance appears at small modulation strength; parametric effects take over at larger modulation amplitudes. Neglecting the direct resonance, the linearized slow dynamics take the form of a Mathieu-type equation with slow Mathieu tongues emerging when modulation frequency $\omega_\mathrm{S} = 2\omega_\mathrm{L}/n$, with $n\in\mathbb{N}$, for large enough  modulation amplitude. As we show later, the relation between the three timescales, set by the modulation rate, relaxation rate and libration rate, gives rise to three qualitatively distinct regimes that involve adiabatic tracking, delayed response, and complex dynamics including period doubling and chaos. 

%%%%%%%%%%%%%%%%%%%%%%%%%%%%%%%%%%%%%%%%%%%%%%%%%%%%%%%%%%%%%%%%%%%%%%%%%%%%%%%%%
%%%%%%%%%%%%%%%%%%%%%%%%%%%%%%%%%%%%%%%%%%%%%%%%%%%%%%%%%%%%%%%%%%%%%%%%%%%%%%%%%

\begin{figure}
    \centering
    \includegraphics[width=\linewidth]{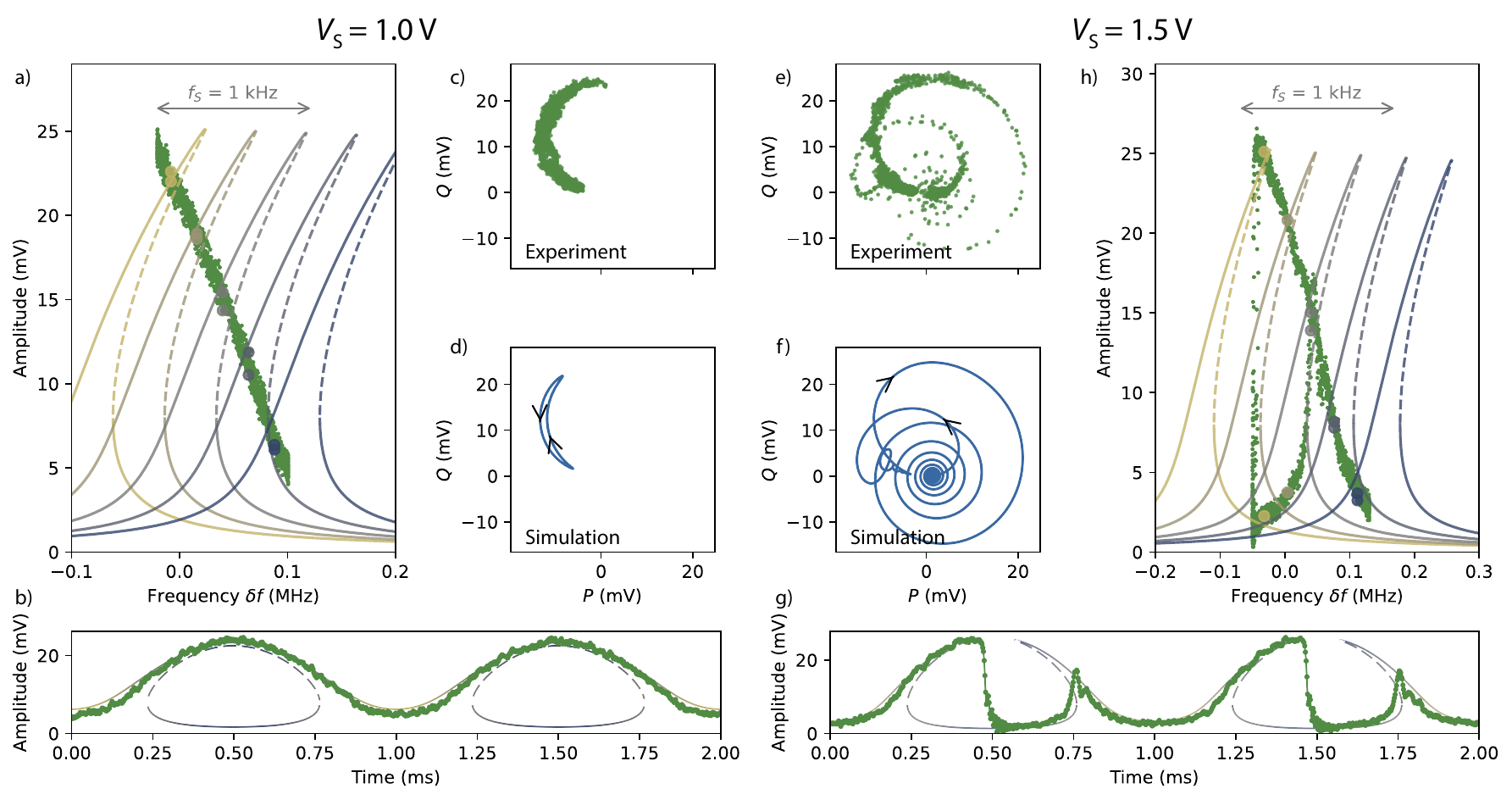}
    \caption{\textbf{Adiabatic tracking under slow tension modulation.}
    \textbf{a--d}, Dynamics at $V_\mathrm{S}=\SI{1.0}{\volt}$, where the resonator follows the instantaneous high-amplitude stable state of the evolving state space.
    \textbf{e--h}, Dynamics at $V_\mathrm{S}=\SI{1.5}{\volt}$, where modulation strength is sufficiently large to reach both saddle-node bifurcations, so the resonator jumps between stable branches periodically. 
    Both measurements are performed at slow modulation frequency $f_\mathrm{S}=\SI{1}{\kilo\hertz}$.
    \textbf{a,h}, Measured response amplitude (green) on the instantaneous nonlinear frequency-response curves.
    The yellow-to-blue curves indicate successive modulation phases; solid and dashed lines denote stable and unstable solutions.
    Coloured circles indicate where the phase of the measured response coincides with that of the same-coloured curves.
    The arrows mark the modulation-induced resonance-frequency excursion.
    The horizontal axis is the effective instantaneous detuning,
    $\delta f(t)=f-2f_0+\hat{f}_0(t)$,
    with $f$ the actual response frequency, $f_0$ the unmodulated resonance frequency and $\hat{f}_0(t)$ its modulation-induced resonance frequency.
    The vertical axis shows the amplitude $\sqrt{P^2+Q^2}$.
    \textbf{b,g}, Measured amplitudes, together with the instantaneous fixed-point amplitudes, over two modulation cycles.
    The responses track the stable branches at $1\,\mathrm{kHz}$.
    \textbf{c,e}, Experimental rotating-frame trajectories in the $(P,Q)$ plane.
    \textbf{d,f}, Corresponding simulations.
    Simulation parameters are obtained from fits to the frequency sweeps, with $A_\mathrm{S}=60\,\mathrm{kHz}$ and $90\,\mathrm{kHz}$ for $V_\mathrm{S}=\SI{1}{\volt}$ and $\SI{1.5}{\volt}$, respectively.}
    \label{fig:adiabatic-tracking}
\end{figure}

\section{Adiabatic tracking, delayed response and period-doubling}

When the fixed points evolve slowly, the resonator can adapt sufficiently quickly to the changing phase-space and remain close to the instantaneous attractors. Conversely, when the equilibria move too rapidly, the resonator can no longer relax towards them in time and develops a delayed response. To explore these limiting regimes experimentally, we drive the graphene resonator with modulation frequencies chosen to lie either well below or well above its intrinsic relaxation rate $f_\mathrm{relax}=\frac{2\Gamma}{2\pi}=\SI{15.6}{\kilo\hertz}$, and compare the resulting responses.

First, we apply a slow modulation at $f_\mathrm{S}=\frac{\omega_\mathrm{S}}{2\pi}=\SI{1}{\kilo\hertz}$, driven at $V_\mathrm{S}=\SI{1}{\volt}$, leading to a regime of quasi-static dynamics. Fig.~\ref{fig:adiabatic-tracking}a showcases the adiabatic tracking of the fixed point. Here, the attractor coincides with the instantaneous high‑amplitude fixed point that, unlike the unstable saddle and the low amplitude attractor, persists throughout the modulation cycle (Fig.~\ref{fig:adiabatic-tracking}b). In the experimental P-Q space of Fig.~\ref{fig:adiabatic-tracking}c and theoretical of Fig.~\ref{fig:adiabatic-tracking}d, the dynamics evolve in a closed orbit according to the periodic modulation of the drum stiffness. This behaviour is visualised in Supplementary Video 1, where the trajectory is animated on the instantaneous phase space. Next, we increase the slow drive voltage $V_\mathrm{S}$ to \SI{1.5}{\volt}, which is sufficiently large for both saddle-node bifurcations to pass the fast drive frequency. As a result, both the high and low amplitude attractors periodically disappear in turn, causing the resonator to switch between them as seen in Fig.~\ref{fig:adiabatic-tracking}g and animated in Supplementary Video 2. Despite this switching, the dynamics remain a closed orbit determined by the periodic stiffness modulation (Fig.~\ref{fig:adiabatic-tracking}e), which now includes two spirals indicating the relaxation towards the attractors. 

\begin{figure}
    \centering
    \includegraphics[width=\linewidth]{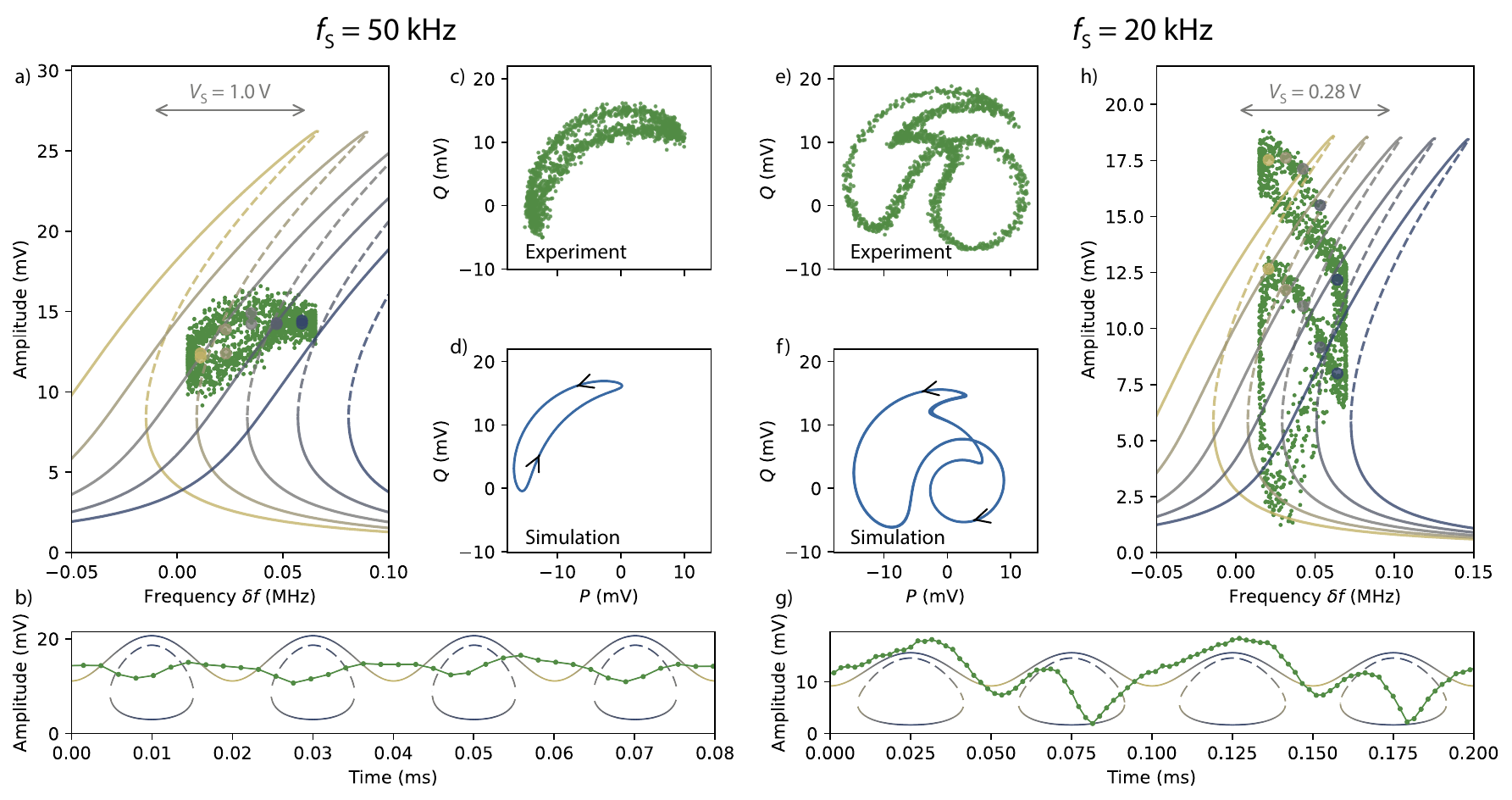}
    \caption{\textbf{Nonadiabatic motion under slow tension modulation.}
    \textbf{a--d}, Dynamics at $f_\mathrm{S}=\SI{50}{\kilo\hertz}$ and $V_\mathrm{S}=\SI{1.0}{\volt}$ ($A_\mathrm{S}=\SI{30}{\kilo\hertz}$), where finite mechanical relaxation produces a pronounced lag.
    \textbf{e--h}, Period-doubled dynamics at $f_\mathrm{S}=\SI{20}{\kilo\hertz}$ and $V_\mathrm{S}=\SI{0.28}{\volt}$ ($A_\mathrm{S}=\SI{27.16}{\kilo\hertz}$).
    \textbf{a,h}, Measured response amplitude (green) on the instantaneous nonlinear frequency-response curves.
    The yellow-to-blue curves indicate successive modulation phases; solid and dashed lines denote stable and unstable solutions.
    Coloured circles indicate where the phase of the measured response coincides with that of the same-coloured curves.
    The horizontal axis is the effective instantaneous detuning,
    $\delta f(t)=f-2f_0+\hat{f}_0(t)$,
    with $f$ the actual response frequency, $f_0$ the unmodulated resonance frequency and $\hat{f}_0(t)$ its modulation-induced resonance frequency.
    The vertical axis shows the amplitude $\sqrt{P^2+Q^2}$.
    \textbf{b,g}, Measured amplitudes together with the instantaneous stable and unstable fixed-point amplitudes over successive modulation cycles.
    For the \SI{50}{\kilo\hertz} case, the response lags behind the stable branches.
    For the \SI{20}{\kilo\hertz} case, the motion repeats only after two modulation periods.
    \textbf{c,e}, Experimental rotating-frame trajectories in the $(P,Q)$ plane.
    \textbf{d,f}, Corresponding simulations, with arrows indicating the direction of motion.
    Simulation parameters are obtained from fits to the frequency sweeps.}
    \label{fig:nonadiabatic-dynamics}
\end{figure}

To explore the opposite limit, we set the slow modulation frequency $f_\mathrm{S}=\SI{50}{\kilo\hertz}$, which is much faster than the relaxation rate. Although the topology of the attractor remains similar to that of the 1 kHz case (Fig.~\ref{fig:nonadiabatic-dynamics}c and d), the fixed points now move more rapidly, and the resonator no longer has sufficient time to relax towards them, resulting in the loss of adiabatic tracking, as shown in Fig.~\ref{fig:nonadiabatic-dynamics}a. Here, the trajectory no longer coincides with the high-amplitude solution (Fig.~\ref{fig:nonadiabatic-dynamics}b). Supplementary Video 3 animates the emergence of this lagged dynamics, as the trajectory attempts to relax towards the fixed point. Additionally, the comparison between the modulation at 1 kHz and 50 kHz reveals insights into the efficacy of the optothermal excitation. Both measurements are performed with $V_\mathrm{S}=\SI{1}{\volt}$; however, the theoretical fit predicts modulation amplitudes of \SI{60}{\kilo\hertz} and \SI{30}{\kilo\hertz} for the low and high $f_\mathrm{S}$, respectively. This reduction in modulation efficiency at higher frequencies is likely associated with the finite thermal response time of the membrane and therefore sets an upper frequency limit for effective optothermal parametric modulation. Together, these measurements establish the two limiting behaviours of the system: adiabatic tracking and delayed response. 

The transition between these limits occurs when the motion of the fixed points becomes comparable to the relaxation rate of the resonator. To probe this regime, we set the modulation frequency to $f_\mathrm{S}=\SI{20}{\kilo\hertz}$, close to the relaxation rate $f_\mathrm{relax}=\SI{15.6}{\kilo\hertz}$. We expect that at this modulation frequency, a slow Mathieu tongue can become accessible. For the high-amplitude fixed point, Eq.~\eqref{eq:slow-flow-eigenfrequency} gives a libration frequency of $f_\mathrm{L}=\omega_\mathrm{L}/(2\pi)=\SI{31.2}{\kilo\hertz}$, so that the condition $\omega_\mathrm{S}\approx 2\omega_\mathrm{L}/3$ is satisfied at $f_\mathrm{S}\approx\SI{20.8}{\kilo\hertz}$.  In fact, as we show in Fig.~\ref{fig:nonadiabatic-dynamics}e-h, for $V_\mathrm{S}=\SI{0.28}{\volt}$, the graphene drum exhibits period-doubled motion. Unlike the adiabatic tracking observed at low $f_\mathrm{S}$ or the delayed response seen at high $f_\mathrm{S}$, Fig.~\ref{fig:nonadiabatic-dynamics}g shows that the resonator requires two modulation periods for the trajectory to repeat, indicating the emergence of a period-two orbit. The origin of this behaviour becomes clear when comparing the trajectory to the evolving fixed points. During one modulation cycle, the response remains associated with the high-amplitude branch. As the low amplitude stable fixed point reappears, the state enters its basin of attraction and begins to orbit around it. When the low amplitude branch subsequently disappears, the trajectory is once again attracted to the high-amplitude branch. Consequently, the resonator alternates between two distinct dynamical cycles, producing a response with twice the period of the modulation (see Supplementary Video 4).  It should be noted that the mechanism behind this physics is distinct from the conventional parametric instabilities previously observed in graphene resonators, where the stiffness is modulated near twice the mechanical resonance frequency $2\omega_0$, typically in the MHz range~\cite{kecskekler2021tuning, jasulaneca2026dynamics}. Here, the graphene drum is instead driven near its fundamental vibrational resonance, while the slow kHz modulation excites parametric-instability tongues associated with the libration frequency $f_\mathrm{L}$ of the rotating-frame.

\begin{figure}
    \centering
    \includegraphics[width=\linewidth]{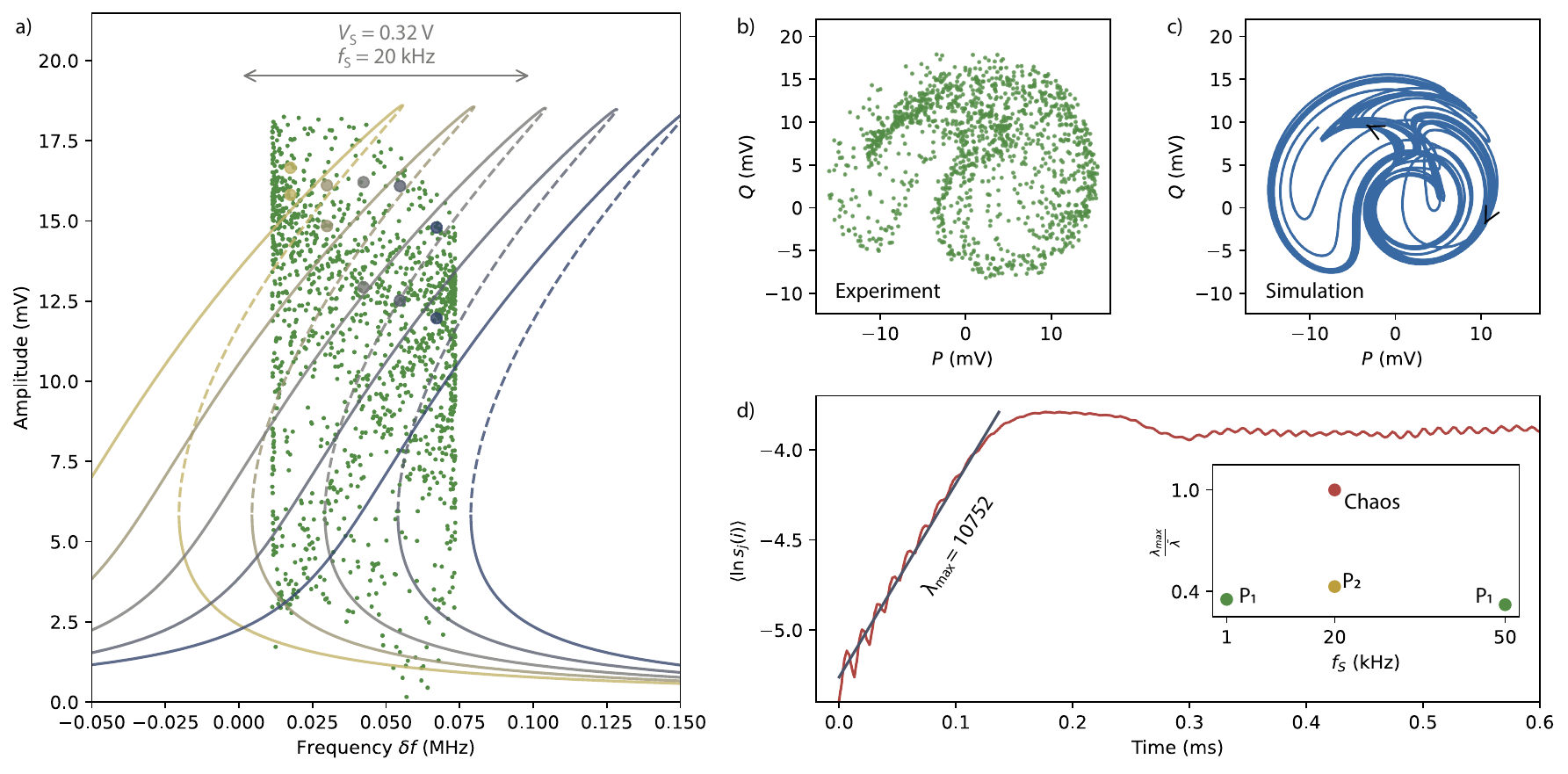}
    \caption{\textbf{Chaotic dynamics as a result of slow tension modulation.}
    \textbf{a}, Measured response amplitude (green) resulting from slow modulation with frequency $f_\mathrm{S}=\SI{20}{\kilo\hertz}$ and amplitude $V_\mathrm{S}=\SI{0.32}{\volt}$ on the instantaneous nonlinear frequency-response curves.
    The yellow-to-blue curves indicate successive modulation phases; solid and dashed lines denote stable and unstable solutions.
    Coloured circles indicate where the phase of the measured response coincides with that of the same-coloured curves.
    The horizontal axis is the effective instantaneous detuning,
    $\delta f(t)=f-2f_0+\hat{f}_0(t)$,
    with $f$ the actual response frequency, $f_0$ the unmodulated resonance frequency and $\hat{f}_0(t)$ its modulation-induced resonance frequency.
    The vertical axis shows the amplitude $\sqrt{P^2+Q^2}$.
    \textbf{b}, Experimental rotating-frame trajectories in the $(P,Q)$ plane.
    \textbf{c}, Corresponding simulations using $A_\mathrm{S}=\SI{31.04}{\kilo\hertz}$, with arrows indicating the direction of motion.
    \textbf{d}, Average separation of the measured response, calculated using Rosenstein method~\cite{Rosenstein1993}. A linear fit, indicated by the black line, is placed on the initial increase to obtain a largest Lyapunov exponent (LLE) $\lambda_\mathrm{max}=\SI{10752}{\per\second}$. The inset compares LLE for data at different $f_\mathrm{S}$ with periodic (P1), period-doubled (P2) or this chaotic motion, normalized relative to the value obtained for chaos ($\bar{\lambda}$). See Supplementary Information~5 for more details.}
    \label{fig:chaos}
\end{figure}

\section{Transition to chaos}

When the phase-space evolution occurs on a timescale comparable to mechanical relaxation, activation of the parametric instability makes the dynamics highly sensitive to the modulation amplitude. As we show next, increasing the voltage applied at the slow tone from \SI{0.28}{\volt}, which showed period-doubling, to \SI{0.32}{\volt} causes a significant change in the dynamics, yielding the emergence of chaotic motion. The resulting dynamics are shown in Fig.~\ref{fig:chaos}a and b. While the trajectory retains a resemblance to the period doubling response seen in Fig.~\ref{fig:nonadiabatic-dynamics}e-h, no two cycles are identical and the periodic switching between the two branches breaks down. The response reveals that the trajectory no longer follows a repeating sequence. Instead, it moves irregularly through phase-space, while being constrained by the evolving attractors and thus reminiscent of chaotic attractors. This behaviour is reproduced via numerical simulations, resulting in similar quadrature spaces in Fig.~\ref{fig:chaos}c. The experimental and numerical trajectories are both animated in Supplementary Video 5.

To prove the emergence of chaotic motion, we calculate the largest Lyapunov exponent (LLE) $\lambda_\mathrm{max}$ from the experimental time series using the Rosenstein method~\cite{Rosenstein1993}. This algorithm computes the separation over time $s_j(t)$ between sets of nearest neighbours $j$, takes the average of the natural logarithm of this separation and fits a line to the initial exponential separation phase to obtain $\lambda_\mathrm{max}$. Fig.~\ref{fig:chaos}d shows the obtained average separation plot for the chaotic trajectory, from which we obtain a positive LLE, confirming exponential sensitivity to initial conditions and therefore chaotic dynamics. We repeat the analysis for different periodic responses, for which we obtain much smaller LLE values as shown in the inset of Fig.~\ref{fig:chaos}d. It should be noted the values are still positive, likely due to noise in the measured data. However, using simulated data we obtain negative LLE for non-chaotic data, as shown with further details in Supplementary Note~5. Additionally, we show the bifurcation diagram and frequency content of the obtained temporal data when driven at $f_\mathrm{S}=\SI{20}{\kilo\hertz}$ in Supplementary Note~6.

\begin{figure}
    \centering
    \includegraphics[width=\linewidth]{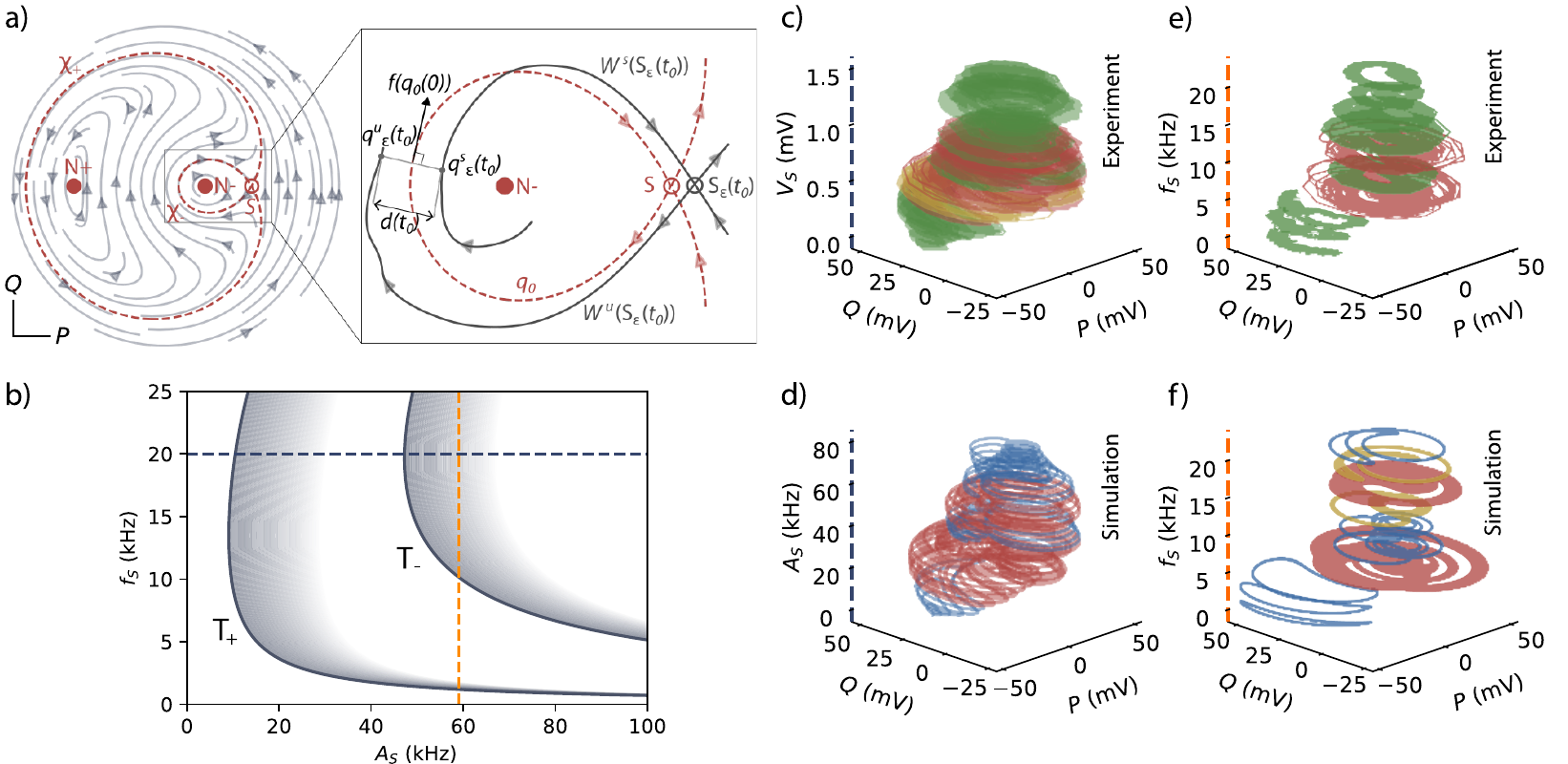}
    \caption{\textbf{Evolution of phase-space structure as a function of frequency and amplitude of slow tension modulation.}
    \textbf{a}, Phase space of the Hamiltonian system   showing the saddle point (S), stable fixed points (N$\pm$), and the corresponding homoclinic orbits ($\chi_\pm$) forming the separatrix. Under perturbation, the stable and unstable manifolds of the saddle, denoted by $W^s$ and $W^u$, respectively, no longer coincide. 
    \textbf{b}, Melnikov prediction of the two instability thresholds $T_\pm$ as a function of modulation amplitude and frequency. The shaded region serves only as a visual guide to the side of the curve where chaos is expected, without implying an additional threshold or boundary. 
    \textbf{c-d}, Experimental and simulated response while sweeping the slow modulation amplitude at constant slow modulation frequency $f_\mathrm{S}=\SI{20}{\kilo\hertz}$. The shaded region projects all responses to the P-Q plane.
    \textbf{e-f}, Experimental and simulated response while sweeping the slow modulation frequency $f_\mathrm{S}$ at constant slow modulation amplitude $A_\mathrm{S}=\SI{59}{\kilo\hertz}$. Since $A_\mathrm{S}$ can not directly be chosen in experiments, $V_\mathrm{S}$ has been swept for each frequency $f_\mathrm{S}$ to obtain the response where $A_\mathrm{S}$ fits closest to \SI{59}{\kilo\hertz}.}
    \label{fig:melnikov}
\end{figure}

The onset of chaos is associated with changes in the phase-space structure governed by the stable and unstable manifolds of the saddle point S. In the unperturbed system, without dissipation or slow modulation, these manifolds coincide and form a separatrix composed of two homoclinic orbits, $\chi_\pm$ (Fig.~\ref{fig:melnikov}a). Each orbit leaves the saddle along $W^u$, returns along $W^s$, and encloses one of the two stable states N$\pm$, thereby separating their respective regions of phase space.

When damping and slow parametric modulation are introduced, the stable and unstable manifolds are perturbed and no longer coincide. As the perturbation evolves, these manifolds can deform and intersect transversely, generating a homoclinic tangle, which is a characteristic geometric signature of chaotic dynamics. The Melnikov method provides a convenient analytical framework for studying this chaos~\cite{guckenheimer2013nonlinear}. 
The Melnikov function is defined as
\begin{equation}
M(t_0) = \int_{-\infty}^{\infty} \mathbf{f}(q_0(t)) \wedge \mathbf{g}(q_0(t), t+t_0)\, dt, 
\label{eq:Melnidef} 
\end{equation}
where $\mathbf{f}$ describes the unperturbed Hamiltonian which for our system is
\begin{equation}
H(P,Q)=-\frac{\hat{\omega}_0^{\,2}-\omega_{\mathrm{F}}^{\,2}}
{4\omega_{\mathrm{F}}}
\left(P^2+Q^2\right)
-\frac{3\gamma}{32\omega_{\mathrm{F}}}
\left(P^2+Q^2\right)^2
+\frac{F_{\mathrm{F}}}{2\omega_{\mathrm{F}}}P.
\label{eq:hamiltonian}
\end{equation} 

The vector $\mathbf{g}$ contains the perturbations due to dissipation and slow parametric modulation (see Supplementary Note~7), and $q_0(t)$ denotes the corresponding unperturbed homoclinic orbit of the saddle S (Fig.~\ref{fig:melnikov}a). The Melnikov function determines, to leading order, the separation between the perturbed stable and unstable manifolds,
\begin{equation}
d(t_0)=q_\varepsilon^u(t_0)-q_\varepsilon^s(t_0)
=\frac{\varepsilon M(t_0)}{|\mathbf{f}(q_0(0))|}+O(\varepsilon^2),
\label{eq:distance}
\end{equation}
where $\varepsilon$ measures the perturbation strength (see Supplementary Note~7). A simple zero of $M(t_0)$ signals a transverse intersection of the manifolds and therefore the onset of a homoclinic tangle and chaotic dynamics~\cite{Melnikov1963}.
The analytical prediction of the two thresholds for the onset of complex dynamics ($T_\pm$) is shown in Fig.~\ref{fig:melnikov}b. 
%where the shaded area represents a visual guide to the region in which chaos is expected to emerge.
The effective Hamiltonian of Eq.~\eqref{eq:hamiltonian} exhibits an asymmetric double-well structure, with the two wells having different energies. Hence the two homoclinic orbits are distinct, resulting in two different Melnikov thresholds, associated with the splitting of the respective homoclinic orbits $\chi_\pm$ under the presence of slow parametric modulation. 
% While the onset of chaos can be obtained using Melnikov theory, the size of the region where chaos appears cannot, which is why the shaded area covers an arbitrary distance. 

To compare theory to experiment, we sweep across the Melnikov thresholds in two separate ways: by sweeping modulation amplitude $A_\mathrm{S}$ at constant frequency $f_\mathrm{S}=\SI{20}{\kilo\hertz}$ and by sweeping $f_\mathrm{S}$ at constant $A_\mathrm{S}=\SI{59}{\kilo\hertz}$, as indicated by the dashed lines in Fig.~\ref{fig:melnikov}b. 
In both cases, we experimentally observe (Fig.~\ref{fig:melnikov}c and e) and numerically predict (Fig.~\ref{fig:melnikov}d and f) two distinct frequency intervals that are signatures of two separate homoclinic instabilities, associated with the sequential onset of transverse intersections between the stable and unstable manifolds that organise the phase-space. To better visualise the changes in response, the evolving P-Q response underlying Fig.~\ref{fig:melnikov}c and d is animated in Supplementary Video~6. Consequently, as $V_\mathrm{S}$ or $f_\mathrm{S}$ is swept, the system first crosses one of the two thresholds, leading to the onset of chaotic dynamics localized in the phase-space region surrounding one stable fixed point, while the other manifold remains regular. As the modulation frequency enters the second instability interval, the corresponding Melnikov function develops simple zeros, indicating the formation of a second homoclinic tangle. 

Although the Melnikov method successfully predicts the existence of two distinct chaotic regions, the analytically derived thresholds deviate slightly from the experimentally observed onsets. Several factors may contribute to this discrepancy. First, the two homoclinic tangles associated with N+ and N- are not dynamically independent and can interact, modifying the local phase-space structure and shifting the onset predicted by the single-orbit Melnikov analysis. Second, persistent chaos is experimentally observed only above a critical forcing amplitude at which the transient chaotic dynamics become sustained. Below this threshold, trajectories can exhibit chaotic motion for a finite time before escaping to a regular attractor.  Above the critical forcing amplitude, this set becomes attracting, leading to persistent chaotic motion \cite{Grebogi1983}. This transition occurs at a forcing amplitude higher than that required for the initial homoclinic manifold intersection predicted by the Melnikov analysis. Third, the Melnikov calculation is based on the conservative unperturbed system, whereas dissipation contracts phase space and alters the stable and unstable manifolds, potentially requiring a stronger modulation amplitude to induce sustained chaotic dynamics.

%%%%%%%%%%%%%%%%%%%%%%%%%%%%%%%%%%%%%%%%%%%%%%%%%%%%%%%%%%%%%%%%%%%%%%%%%%%%%%%%%
%%%%%%%%%%%%%%%%%%%%%%%%%%%%%%%%%%%%%%%%%%%%%%%%%%%%%%%%%%%%%%%%%%%%%%%%%%%%%%%%%

\section{Discussion}

In this work, we experimentally realize a nonlinear graphene resonator whose rotating-frame phase-space structure is continuously reshaped by slow optothermal tension modulation. The resonant drive establishes a bistable response, while the parametric modulation moves the stable and unstable fixed points that organize the dynamics. At low modulation rates, the response adiabatically follows these evolving states; as the modulation rate approaches the mechanical relaxation rate, adiabatic following can no longer be maintained and parametric instabilities emerge in the slow dynamics. These instabilities occur near $2\omega_L/n$, where $\omega_L$ is the libration frequency around the rotating-frame fixed points, in contrast to the conventional Mathieu tongues at $2\omega_0/n$ set by the mechanical resonance frequency~\cite{nayfeh_nonlinear_1979,Dolleman2018,Kartal2026stability}. In our simulations and experiments, the second- to fifth-order instabilities are activated, whereas the primary instability is not observed, likely because it lies at modulation frequencies well above the mechanical relaxation rate (Supplementary Note~4). Increasing the modulation amplitude in this regime leads to period doubling and chaos, with similar behaviour reproduced in two additional graphene drums (Supplementary Note~8). Moreover, the route to this complex dynamics differs from previous approaches in which the nonlinear response was controlled by modulating the external forcing~\cite{houri2020chaos,Defoort2021,Catalini2025,Kumar2026,Eshel2026twotone}. Here, the resonant drive establishes the nonlinear response, whereas modulation of the membrane tension shifts the resonance frequency and thereby controls the motion of the rotating-frame fixed points. These two ingredients can be tuned independently, providing direct access to the regime in which phase-space evolution and mechanical relaxation occur on comparable timescales. More broadly, our results show that parametric modulation can destabilize not only the fast mechanical oscillation, but also the slower rotating-frame dynamics, providing direct control over the transition from adiabatic to nonadiabatic and chaotic motion in a single nonlinear nanomechanical mode.

The small footprint of graphene drums further makes this approach attractive for parallel implementations. Arrays containing many individually addressable resonators could, in principle, be operated at different points across the adiabatic-to-chaotic transition, providing a spatially distributed set of nonlinear dynamical states. Such an architecture may be relevant to physical reservoir computing, where operation near the edge of chaos has been associated with enhanced computational performance~\cite{Carroll2020RcChaos,Pathak2018Prediction,Cucchi2022reservoir}. Likewise, the strong sensitivity of the response to slow perturbations may enable sensing based on stochastic resonance~\cite{Gammaitoni1998stochastic,Roldan2024stochastic}. More generally, dynamically controllable phase-spaces could provide a route to adaptive nonlinear devices whose operating regime can be controlled through low-frequency parameter modulation. Beyond nanomechanical resonators, we expect this type of nonadiabatic dynamics to also appear in other driven-dissipative nonlinear systems, such as superconducting Kerr oscillators and flux-tunable Josephson bifurcation amplifiers \cite{ameye2026fewphoton,Vijay2009josephson}, where flux modulation would play the role of tension modulation. Similar phase-space structures arise in modulated cold-atom traps \cite{kim2006spontaneous,Hensinger2001dynamical}, in which slow trap modulation would have the same effect. Finally, the parameter space explored here represents only part of the broader slow parametric-fast direct paradigm. In addition to the slow modulation frequency and amplitude, the fast drive plays a large role in determining the underlying phase-space and therefore the accessibility of complex dynamics. In particular, the distance between the fast drive frequency and the saddle-node bifurcations could strongly influence the emergence of period doubling and chaos.  The combination of experimentally controllable phase-space evolution and tunable proximity to bifurcation points therefore offers a promising platform for future studies of this phenomenon.

%%%%%%%%%%%%%%%%%%%%%%%%%%%%%%%%%%%%%%%%%%%%%%%%%%%%%%%%%%%%%%%%%%%%%%%%%%%%%%%%%
%%%%%%%%%%%%%%%%%%%%%%%%%%%%%%%%%%%%%%%%%%%%%%%%%%%%%%%%%%%%%%%%%%%%%%%%%%%%%%%%%

%%%%%%%%%%%%%%%%%%%%%%%%%%%%%%%%%%%%%%%%%%%%%%%%%%%%%%%%%%%%%%%%%%%%%%%%%%%%%%%%%
%%%%%%%%%%%%%%%%%%%%%%%%%%%%%%%%%%%%%%%%%%%%%%%%%%%%%%%%%%%%%%%%%%%%%%%%%%%%%%%%%

\section{Methods}
\subsection{Laser interferometry}
Graphene drums are actuated and detected through a single optical port using the interferometric setup shown in Fig.~\ref{fig:physics-and-explanation}b.  A blue diode laser ($\lambda = \SI{488}{\nano\meter}$) provides optothermal actuation, by modulating the laser intensity. The membrane and the substrate form a Fabry--P\'erot cavity, such that membrane displacement modulates the intensity of the red, reflected helium--neon laser ($\lambda = \SI{632.8}{\nano\meter}$), which is measured by a photodiode. The blue and red laser paths are combined by a dichroic mirror, and a polarizing beam splitter and quarter-wave plate are used to separate the incoming from the reflected red laser and redirect the reflected light onto the photodiode. Signal generation for actuation and signal analysis of the photodiode output are both done using a Zurich Instruments HF2LI lock-in amplifier.

\subsection{Measurement protocol}
\label{sec:protocol}
The blue laser intensity is modulated by the sum of two harmonic tones as $V_\mathrm{S}\cos(\omega_\mathrm{S}t)+V_\mathrm{F}\cos(\omega_\mathrm{F}t)$, generated by a lock-in amplifier. The photodiode signal is demodulated at the fast drive frequency $f_\mathrm{F}$ by the same lock-in amplifier to directly obtain the in-phase and out-of-phase quadratures ($P, Q$). Throughout the measurement, amplitudes are recorded as photodetector voltages. All measurements are performed at room temperature in a vacuum chamber at a pressure of \SI{8e-6}{\milli\bar}. 

Before each measurement, a sweep of $f_\mathrm{F}$ without the slow drive is performed. This is done to combat a drift in resonance frequency and therefore ensure the exact experimental conditions are known for each dataset. The sweep is performed at the same drive voltage $V_\mathrm{F}$ as the following measurement. The frequency is swept across 400 points, at a demodulation bandwidth of \SI{100}{\kilo\hertz}. The resulting Duffing response from this sweep is fitted to an analytical model to extract the resonance frequency $f_0$ (see Supplementary Note~1 for the fitting procedure). From the fitted resonance frequency, the fast drive frequency is defined as $f_\mathrm{F}=f_0+
\SI{45}{\kilo\hertz}$ to place it in the bistable regime. 

After defining the fast drive frequency $f_\mathrm{F}$, we start the measurement by activating the fast tone in order to establish the resonant, nonlinear oscillations. The frequency $f_\mathrm{F}$ is swept up from the previously identified resonance frequency to the desired fast frequency to ensure the initial conditions of the sweep are on the high branch of the Duffing response. Alternatively, $f_\mathrm{F}$ could be kept constant while $V_\mathrm{F}$ is swept up to reach the low branch. 

Next, the slow tone is generated at frequency $f_\mathrm{S}$ and drive amplitude $V_\mathrm{S}$ and superimposed on the fast tone. The response is measured for \SIrange[range-phrase=--,range-units=single]{3}{5}{\milli\second}, at a bandwidth of \SI{100}{\kilo\hertz} and acquisition rate of \SI{460}{\kilo\hertz}.

\section{Author contributions}

P.B. and F.A. conceived the idea. H.L. and C.F.D.W. performed the experiments; Simulations were done by C.F.D.W. and H.L.; P.B. further developed the analytical model to analyse chaos. Data analysis and interpretation was done by C.F.D.W., H.L., P.B., J.E.S., P.G.S. and F.A.; The manuscript was written by C.F.D.W.,  P.B. and F.A., with input from all authors; The project was supervised by F.A.

\section{Acknowledgements}

Funded by the European Union (ERC Consolidator, NCANTO, 101125458). Views and opinions expressed are, however, those of the author(s) only and do not necessarily reflect those of the European Union or the European Research Council. Neither the European Union nor the granting authority can be held responsible for them. The authors thank SoundCell B.V. for providing the graphene samples. C.F.D.W. acknowledges the fruitful discussions with Dr. Alberto Martin-Perez, Dr. Zichao Li, Mrs. Enise Kartal and Mr. Kushal Swamy. F.A., C.F.D.W. and P.B. acknowledge Dr. Robin Dolleman for exploring this concept in the initial stages of this project.

\bibliography{biblio}

\begin{thebibliography}{10}
\expandafter\ifx\csname url\endcsname\relax
  \def\url#1{\burl{#1}}\fi
\expandafter\ifx\csname urlprefix\endcsname\relax\def\urlprefix{URL }\fi
\providecommand{\bibinfo}[2]{#2}
\providecommand{\eprint}[2][]{\url{#2}}
\providecommand{\doi}[1]{\url{https://doi.org/#1}}
\bibcommenthead

\bibitem{bachtold2022mesoscopic}
\bibinfo{author}{Bachtold, A.}, \bibinfo{author}{Moser, J.} \& \bibinfo{author}{Dykman, M.}
\newblock \bibinfo{title}{Mesoscopic physics of nanomechanical systems}.
\newblock \emph{\bibinfo{journal}{Reviews of Modern Physics}} \textbf{\bibinfo{volume}{94}}, \bibinfo{pages}{045005} (\bibinfo{year}{2022}).

\bibitem{eichler2023classical}
\bibinfo{author}{Eichler, A.} \& \bibinfo{author}{Zilberberg, O.}
\newblock \emph{\bibinfo{title}{Classical and Quantum Parametric Phenomena}}  (\bibinfo{publisher}{Oxford University Press}, \bibinfo{address}{Oxford}, \bibinfo{year}{2023}).

\bibitem{strogatz2024nonlinear}
\bibinfo{author}{Strogatz, S.~H.}
\newblock \emph{\bibinfo{title}{Nonlinear dynamics and chaos: with applications to physics, \\biology, chemistry, and engineering}}  (\bibinfo{publisher}{CRC press}, \bibinfo{year}{2024}).

\bibitem{Huber2020Spectral}
\bibinfo{author}{Huber, J.~S.} \emph{et~al.}
\newblock \bibinfo{title}{Spectral evidence of squeezing of a weakly damped driven nanomechanical mode}.
\newblock \emph{\bibinfo{journal}{Phys. Rev. X}} \textbf{\bibinfo{volume}{10}}, \bibinfo{pages}{021066} (\bibinfo{year}{2020}).

\bibitem{matheny2014phase}
\bibinfo{author}{Matheny, M.~H.} \emph{et~al.}
\newblock \bibinfo{title}{Phase synchronization of two anharmonic nanomechanical oscillators}.
\newblock \emph{\bibinfo{journal}{Physical review letters}} \textbf{\bibinfo{volume}{112}}, \bibinfo{pages}{014101} (\bibinfo{year}{2014}).

\bibitem{houri2026three}
\bibinfo{author}{Houri, S.}, \bibinfo{author}{Asano, M.}, \bibinfo{author}{Okamoto, H.} \& \bibinfo{author}{Yamaguchi, H.}
\newblock \bibinfo{title}{A three-modes kuramoto network in a single nonlinear microelectromechanical device}.
\newblock \emph{\bibinfo{journal}{Applied Physics Letters}} \textbf{\bibinfo{volume}{129}} (\bibinfo{year}{2026}).

\bibitem{kecskekler2021tuning}
\bibinfo{author}{Ke{\c{s}}kekler, A.} \emph{et~al.}
\newblock \bibinfo{title}{Tuning nonlinear damping in graphene nanoresonators by parametric--direct internal resonance}.
\newblock \emph{\bibinfo{journal}{Nature communications}} \textbf{\bibinfo{volume}{12}}, \bibinfo{pages}{1099} (\bibinfo{year}{2021}).

\bibitem{shoshani2021resonant}
\bibinfo{author}{Shoshani, O.} \& \bibinfo{author}{Shaw, S.~W.}
\newblock \bibinfo{title}{Resonant modal interactions in micro/nano-mechanical structures}.
\newblock \emph{\bibinfo{journal}{Nonlinear Dynamics}} \textbf{\bibinfo{volume}{104}}, \bibinfo{pages}{1801--1828} (\bibinfo{year}{2021}).

\bibitem{yan2022energy}
\bibinfo{author}{Yan, Y.}, \bibinfo{author}{Dong, X.}, \bibinfo{author}{Huang, L.}, \bibinfo{author}{Moskovtsev, K.} \& \bibinfo{author}{Chan, H.~B.}
\newblock \bibinfo{title}{Energy transfer into period-tripled states in coupled electromechanical modes at internal resonance}.
\newblock \emph{\bibinfo{journal}{Phys. Rev. X}} \textbf{\bibinfo{volume}{12}}, \bibinfo{pages}{031003} (\bibinfo{year}{2022}).

\bibitem{Belardinelli2025}
\bibinfo{author}{Belardinelli, P.}, \bibinfo{author}{Yang, W.}, \bibinfo{author}{Bachtold, A.}, \bibinfo{author}{Dykman, M.~I.} \& \bibinfo{author}{Alijani, F.}
\newblock \bibinfo{title}{Hidden vibrational bistability revealed by intrinsic fluctuations of a carbon nanotube}.
\newblock \emph{\bibinfo{journal}{Nano Letters}} \textbf{\bibinfo{volume}{25}}, \bibinfo{pages}{8443--8449} (\bibinfo{year}{2025}).

\bibitem{boness2026nonequilibrium}
\bibinfo{author}{Bone{\ss}, D.~K.}, \bibinfo{author}{Margiani, G.}, \bibinfo{author}{Belzig, W.}, \bibinfo{author}{Eichler, A.} \& \bibinfo{author}{Zilberberg, O.}
\newblock \bibinfo{title}{Nonequilibrium kramers turnover in a kerr parametric oscillator}.
\newblock \emph{\bibinfo{journal}{arXiv preprint arXiv:2604.19527}}  (\bibinfo{year}{2026}).

\bibitem{czaplewski2018bifurcation}
\bibinfo{author}{Czaplewski, D.~A.} \emph{et~al.}
\newblock \bibinfo{title}{Bifurcation generated mechanical frequency comb}.
\newblock \emph{\bibinfo{journal}{Physical review letters}} \textbf{\bibinfo{volume}{121}}, \bibinfo{pages}{244302} (\bibinfo{year}{2018}).

\bibitem{Ochs2022FrequencyComb}
\bibinfo{author}{Ochs, J.~S.} \emph{et~al.}
\newblock \bibinfo{title}{Frequency comb from a single driven nonlinear nanomechanical mode}.
\newblock \emph{\bibinfo{journal}{Phys. Rev. X}} \textbf{\bibinfo{volume}{12}}, \bibinfo{pages}{041019} (\bibinfo{year}{2022}).

\bibitem{Catalini2025}
\bibinfo{author}{Catalini, L.} \emph{et~al.}
\newblock \bibinfo{title}{Slow and fast topological dynamical phase transitions in a duffing resonator driven by two detuned tones}.
\newblock \emph{\bibinfo{journal}{Phys. Rev. Res.}} \textbf{\bibinfo{volume}{7}}, \bibinfo{pages}{033058} (\bibinfo{year}{2025}).

\bibitem{Kumar2026}
\bibinfo{author}{Kumar, S.~S.}, \bibinfo{author}{del Pino, J.}, \bibinfo{author}{Catalini, L.}, \bibinfo{author}{Eichler, A.} \& \bibinfo{author}{Zilberberg, O.}
\newblock \bibinfo{title}{Dynamical phase transitions across slow and fast regimes in a two-tone driven duffing resonator}.
\newblock \emph{\bibinfo{journal}{Phys. Rev. Res.}} \textbf{\bibinfo{volume}{8}}, \bibinfo{pages}{023298} (\bibinfo{year}{2026}).

\bibitem{Eshel2026twotone}
\bibinfo{author}{Eshel, A.}, \bibinfo{author}{Choi, H.}, \bibinfo{author}{Choi, H.}, \bibinfo{author}{Shaw, S.~W.} \& \bibinfo{author}{Shoshani, O.}
\newblock \bibinfo{title}{Nonlinear dynamics of a duffing resonator driven by two near-resonant tones}.
\newblock \emph{\bibinfo{journal}{Journal of Sound and Vibration}} \textbf{\bibinfo{volume}{643}}, \bibinfo{pages}{120020} (\bibinfo{year}{2026}).

\bibitem{ribeiro2013coherent}
\bibinfo{author}{Ribeiro, H.}, \bibinfo{author}{Burkard, G.}, \bibinfo{author}{Petta, J.~R.}, \bibinfo{author}{Lu, H.} \& \bibinfo{author}{Gossard, A.~C.}
\newblock \bibinfo{title}{Coherent adiabatic spin control in the presence of charge noise using tailored pulses}.
\newblock \emph{\bibinfo{journal}{Physical Review Letters}} \textbf{\bibinfo{volume}{110}}, \bibinfo{pages}{086804} (\bibinfo{year}{2013}).

\bibitem{quintana2013cavity}
\bibinfo{author}{Quintana, C.} \emph{et~al.}
\newblock \bibinfo{title}{Cavity-mediated entanglement generation via landau-zener interferometry}.
\newblock \emph{\bibinfo{journal}{Physical review letters}} \textbf{\bibinfo{volume}{110}}, \bibinfo{pages}{173603} (\bibinfo{year}{2013}).

\bibitem{faust2012nonadiabatic}
\bibinfo{author}{Faust, T.} \emph{et~al.}
\newblock \bibinfo{title}{Nonadiabatic dynamics of two strongly coupled nanomechanical resonator modes}.
\newblock \emph{\bibinfo{journal}{Physical review letters}} \textbf{\bibinfo{volume}{109}}, \bibinfo{pages}{037205} (\bibinfo{year}{2012}).

\bibitem{yanchuk2026singular}
\bibinfo{author}{Yanchuk, S.}, \bibinfo{author}{Wieczorek, S.}, \bibinfo{author}{Jard{\~A}{\`g}n-Kojakhmetov, H.} \& \bibinfo{author}{Alkhayuon, H.}
\newblock \bibinfo{title}{Singular basins in multiscale systems: tunneling between stable states}.
\newblock \emph{\bibinfo{journal}{arXiv preprint arXiv:2601.02001}}  (\bibinfo{year}{2026}).

\bibitem{Dolleman2018}
\bibinfo{author}{Dolleman, R.~J.} \emph{et~al.}
\newblock \bibinfo{title}{Opto-thermally excited multimode parametric resonance in graphene membranes}.
\newblock \emph{\bibinfo{journal}{Scientific Reports}} \textbf{\bibinfo{volume}{8}}, \bibinfo{pages}{9366} (\bibinfo{year}{2018}).

\bibitem{Kartal2026stability}
\bibinfo{author}{Kartal, E.} \emph{et~al.}
\newblock \bibinfo{title}{Frequency stability of graphene nonlinear parametric oscillator}.
\newblock \emph{\bibinfo{journal}{Nano Letters}} \textbf{\bibinfo{volume}{26}}, \bibinfo{pages}{7501--7508} (\bibinfo{year}{2026}).

\bibitem{rathi2026tunablenonlinearlandscapesgraphene}
\bibinfo{author}{Rathi, A.~K.} \emph{et~al.}
\newblock \bibinfo{title}{Tunable nonlinear landscapes in graphene nanoelectromechanical systems} (\bibinfo{year}{2026}).

\bibitem{steeneken2021dynamics}
\bibinfo{author}{Steeneken, P.~G.}, \bibinfo{author}{Dolleman, R.~J.}, \bibinfo{author}{Davidovikj, D.}, \bibinfo{author}{Alijani, F.} \& \bibinfo{author}{van~der Zant, H. S.~J.}
\newblock \bibinfo{title}{Dynamics of 2d material membranes}.
\newblock \emph{\bibinfo{journal}{2D Materials}} \textbf{\bibinfo{volume}{8}}, \bibinfo{pages}{042001} (\bibinfo{year}{2021}).

\bibitem{davidovikj2017nonlinear}
\bibinfo{author}{Davidovikj, D.} \emph{et~al.}
\newblock \bibinfo{title}{Nonlinear dynamic characterization of two-dimensional materials}.
\newblock \emph{\bibinfo{journal}{Nature Communications}} \textbf{\bibinfo{volume}{8}}, \bibinfo{pages}{1253} (\bibinfo{year}{2017}).

\bibitem{kryloff1947introduction}
\bibinfo{author}{Kryloff, N.}, \bibinfo{author}{Bogoliubov, N.} \& \bibinfo{author}{Lefschetz, S.}
\newblock \emph{\bibinfo{title}{Introduction to Non-linear Mechanics}} Annals of mathematics studies (\bibinfo{publisher}{Princeton University Press}, \bibinfo{year}{1947}).

\bibitem{jasulaneca2026dynamics}
\bibinfo{author}{Jasulaneca, L.} \emph{et~al.}
\newblock \bibinfo{title}{Dynamics of amorphous membranes in the two-dimensional limit}.
\newblock \emph{\bibinfo{journal}{arXiv preprint arXiv:2608.10486}}  (\bibinfo{year}{2026}).

\bibitem{Rosenstein1993}
\bibinfo{author}{Rosenstein, M.~T.}, \bibinfo{author}{Collins, J.~J.} \& \bibinfo{author}{De~Luca, C.~J.}
\newblock \bibinfo{title}{A practical method for calculating largest lyapunov exponents from small data sets}.
\newblock \emph{\bibinfo{journal}{Physica D: Nonlinear Phenomena}} \textbf{\bibinfo{volume}{65}}, \bibinfo{pages}{117--134} (\bibinfo{year}{1993}).

\bibitem{guckenheimer2013nonlinear}
\bibinfo{author}{Guckenheimer, J.} \& \bibinfo{author}{Holmes, P.}
\newblock \emph{\bibinfo{title}{Nonlinear Oscillations, Dynamical Systems, and \\Bifurcations of Vector Fields}} Applied Mathematical Sciences (\bibinfo{publisher}{Springer}, \bibinfo{address}{New York}, \bibinfo{year}{2013}).

\bibitem{Melnikov1963}
\bibinfo{author}{Melnikov, V.}
\newblock \bibinfo{title}{On the stability of a center for time-periodic perturbations}.
\newblock \emph{\bibinfo{journal}{Transaction of the Moskow Mathematical Society}} \textbf{\bibinfo{volume}{12}}, \bibinfo{pages}{3--52} (\bibinfo{year}{1963}).

\bibitem{Grebogi1983}
\bibinfo{author}{Grebogi, C.}, \bibinfo{author}{Ott, E.} \& \bibinfo{author}{Yorke, J.~A.}
\newblock \bibinfo{title}{Crises, sudden changes in chaotic attractors, and transient chaos}.
\newblock \emph{\bibinfo{journal}{Physica D: Nonlinear Phenomena}} \textbf{\bibinfo{volume}{7}}, \bibinfo{pages}{181--200} (\bibinfo{year}{1983}).

\bibitem{nayfeh_nonlinear_1979}
\bibinfo{author}{Nayfeh, A.~H.} \& \bibinfo{author}{Mook, D.~T.}
\newblock \emph{\bibinfo{title}{Nonlinear oscillations}} Pure and applied mathematics (\bibinfo{publisher}{Wiley}, \bibinfo{address}{New York}, \bibinfo{year}{1979}).

\bibitem{houri2020chaos}
\bibinfo{author}{Houri, S.} \emph{et~al.}
\newblock \bibinfo{title}{Generic rotating-frame-based approach to chaos generation in nonlinear micro- and nanoelectromechanical system resonators}.
\newblock \emph{\bibinfo{journal}{Phys. Rev. Lett.}} \textbf{\bibinfo{volume}{125}}, \bibinfo{pages}{174301} (\bibinfo{year}{2020}).

\bibitem{Defoort2021}
\bibinfo{author}{Defoort, M.}, \bibinfo{author}{Rufer, L.}, \bibinfo{author}{Fesquet, L.} \& \bibinfo{author}{Basrour, S.}
\newblock \bibinfo{title}{A dynamical approach to generate chaos in a micromechanical resonator}.
\newblock \emph{\bibinfo{journal}{Microsystems {\&} Nanoengineering}} \textbf{\bibinfo{volume}{7}}, \bibinfo{pages}{17} (\bibinfo{year}{2021}).

\bibitem{Carroll2020RcChaos}
\bibinfo{author}{Carroll, T.~L.}
\newblock \bibinfo{title}{Do reservoir computers work best at the edge of chaos?}
\newblock \emph{\bibinfo{journal}{Chaos: An Interdisciplinary Journal of Nonlinear Science}} \textbf{\bibinfo{volume}{30}}, \bibinfo{pages}{121109} (\bibinfo{year}{2020}).

\bibitem{Pathak2018Prediction}
\bibinfo{author}{Pathak, J.}, \bibinfo{author}{Hunt, B.}, \bibinfo{author}{Girvan, M.}, \bibinfo{author}{Lu, Z.} \& \bibinfo{author}{Ott, E.}
\newblock \bibinfo{title}{Model-free prediction of large spatiotemporally chaotic systems from data: A reservoir computing approach}.
\newblock \emph{\bibinfo{journal}{Phys. Rev. Lett.}} \textbf{\bibinfo{volume}{120}}, \bibinfo{pages}{024102} (\bibinfo{year}{2018}).

\bibitem{Cucchi2022reservoir}
\bibinfo{author}{Cucchi, M.}, \bibinfo{author}{Abreu, S.}, \bibinfo{author}{Ciccone, G.}, \bibinfo{author}{Brunner, D.} \& \bibinfo{author}{Kleemann, H.}
\newblock \bibinfo{title}{Hands-on reservoir computing: a tutorial for practical implementation}.
\newblock \emph{\bibinfo{journal}{Neuromorphic Computing and Engineering}} \textbf{\bibinfo{volume}{2}}, \bibinfo{pages}{032002} (\bibinfo{year}{2022}).

\bibitem{Gammaitoni1998stochastic}
\bibinfo{author}{Gammaitoni, L.}, \bibinfo{author}{H\"anggi, P.}, \bibinfo{author}{Jung, P.} \& \bibinfo{author}{Marchesoni, F.}
\newblock \bibinfo{title}{Stochastic resonance}.
\newblock \emph{\bibinfo{journal}{Rev. Mod. Phys.}} \textbf{\bibinfo{volume}{70}}, \bibinfo{pages}{223--287} (\bibinfo{year}{1998}).

\bibitem{Roldan2024stochastic}
\bibinfo{author}{Roldán, J.~B.} \emph{et~al.}
\newblock \bibinfo{title}{Stochastic resonance in 2d materials based memristors}.
\newblock \emph{\bibinfo{journal}{npj 2D Materials and Applications}} \textbf{\bibinfo{volume}{8}} (\bibinfo{year}{2024}).

\bibitem{ameye2026fewphoton}
\bibinfo{author}{Ameye, O.}, \bibinfo{author}{Koenig, J.}, \bibinfo{author}{Potts, C.}, \bibinfo{author}{Zilberberg, O.} \& \bibinfo{author}{Steele, G.}
\newblock \bibinfo{title}{Few-photon degenerate parametric resonance in a two-tone driven microwave resonator} (\bibinfo{year}{2026}).

\bibitem{Vijay2009josephson}
\bibinfo{author}{Vijay, R.}, \bibinfo{author}{Devoret, M.~H.} \& \bibinfo{author}{Siddiqi, I.}
\newblock \bibinfo{title}{Invited review article: The josephson bifurcation amplifier}.
\newblock \emph{\bibinfo{journal}{Review of Scientific Instruments}} \textbf{\bibinfo{volume}{80}}, \bibinfo{pages}{111101} (\bibinfo{year}{2009}).
\newblock \urlprefix\url{https://doi.org/10.1063/1.3224703}.

\bibitem{kim2006spontaneous}
\bibinfo{author}{Kim, K.} \emph{et~al.}
\newblock \bibinfo{title}{Spontaneous symmetry breaking of population in a nonadiabatically driven atomic trap: An ising-class phase transition}.
\newblock \emph{\bibinfo{journal}{Phys. Rev. Lett.}} \textbf{\bibinfo{volume}{96}}, \bibinfo{pages}{150601} (\bibinfo{year}{2006}).
\newblock \urlprefix\url{https://link.aps.org/doi/10.1103/PhysRevLett.96.150601}.

\bibitem{Hensinger2001dynamical}
\bibinfo{author}{Hensinger, W.~K.} \emph{et~al.}
\newblock \bibinfo{title}{Dynamical tunnelling of ultracold atoms}.
\newblock \emph{\bibinfo{journal}{Nature}} \textbf{\bibinfo{volume}{412}}, \bibinfo{pages}{52--55} (\bibinfo{year}{2001}).
\newblock \urlprefix\url{https://doi.org}.

\end{thebibliography}

\clearpage



\section*{Supplementary material}

# Supplementary Information: Nonadiabatic dynamics under slow parametric modulation in nonlinear graphene resonators

Chris F. D. Wattjes[1*], Haemin Lim[1], Pierpaolo Belardinelli[2], John E. Sader[3], Peter G. Steeneken[1], Farbod Alijani[1*]

[1]Department of Precision and Microsystems Engineering, Delft University of Technology, Mekelweg 2, Delft, 2628 CD, The Netherlands.
[2]Department of Construction, Civil Engineering and Architecture, Polytechnic University of Marche, via Brecce Bianche, Ancona, 60131, Italy.
[3]Lynn Booth & Kent Kresa Department of Aerospace & Department of Applied Physics and Materials Science, California Institute of Technology, Pasadena, CA 91125, USA.

*Corresponding author(s). E-mail(s): C.F.D.Wattjes-1@tudelft.nl; F.Alijani@tudelft.nl;

# 1 Nonlinear parameter estimation

In order to extract linear and nonlinear parameters from the Duffing curves, we numerically fit the curves with the model

$$\ddot{x} + 2\Gamma\dot{x} + \omega_0^2 x + \gamma x^3 = F_{\mathrm{F}} \cos(\omega_{\mathrm{F}} t), \tag{1}$$

which is Eq. 1 from the main text in the absence of slow drive.

The fitting procedure is applied to normalized data, with frequency being normalized w.r.t. the mean sweep-frequency $f_{\mathrm{mean}} = \frac{1}{2}(f_{\mathrm{min}} + f_{\mathrm{max}})$ and amplitude w.r.t. the maximum experimental amplitude $A_{\mathrm{max}}$.

As a first step, we fit the backbone curve to the high-amplitude ($A > \frac{1}{2}A_{\mathrm{max}}$) data of the sweep. With the backbone curve defined as

$$f^2 = f_0^2 + \frac{3\gamma}{16\pi^2}A^2, \tag{2}$$

we fit a line of the form $y = \gamma x + f_0^2$, with $y = f^2$ and $x = \frac{3}{16\pi^2}A^2$ to obtain the Duffing constant and resonance frequency.

Next, we utilize the off-resonance response to get an initial estimate for the damping constant. After excluding the high-amplitude ($A > \frac{1}{2}A_{\mathrm{max}}$) points used for the backbone fit, we retain, on each side of the resonance, only the outer half of the remaining points, meaning those farthest from resonance in frequency, as the off-resonance data set. To this data, we fit a linear response function

$$A = \frac{F_{\mathrm{F}}}{\sqrt{(\omega_0^2 - \omega^2)^2 + (2\Gamma\omega\omega_0)^2}} = \frac{\Gamma f_0 A_{max}}{\sqrt{\pi^2\left(f_0^2 - f^2\right)^2 + \left(\Gamma f_0\right)^2}}, \tag{3}$$

where the drive force is related to the peak amplitude as $F_{\mathrm{F}} = 2\Gamma\omega_0 A_{max}$ and $f_0$ is defined as its backbone estimate, to obtain $\Gamma$.

Finally, we apply harmonic balancing method to Eq. (1) by substituting $x = A\cos(\omega_{\mathrm{F}})$, yielding a polynomial for $A$ in terms of $\omega_0$, $\Gamma$, $\gamma$ and $F_{\mathrm{F}}$, which we numerically solve for its roots. When multiple real, positive roots are found, as in the bistable regime, we retain the largest amplitude solution, as should be the case in the experiments. Based on this, we optimize $\omega_0$, $\Gamma$, $\gamma$ and $F_{\mathrm{F}}$ by minimizing the least squares error between the analytically and experimentally obtained amplitudes. Here, the parameters obtained in the first two steps are used as initial values. The final and intermediate results for the data seen in Fig. 1c of the main text are shown in Fig. 1.

# 2 Scaling of optothermal forcing and parametric modulation

To determine how the different components of the optothermal actuation scale in the nonlinear membrane dynamics, we consider a circular membrane with a small out-of-plane offset and a temperature-dependent in-plane strain. To that end, we approximate the transverse displacement using the fundamental axisymmetric mode of the circular membrane as follows,

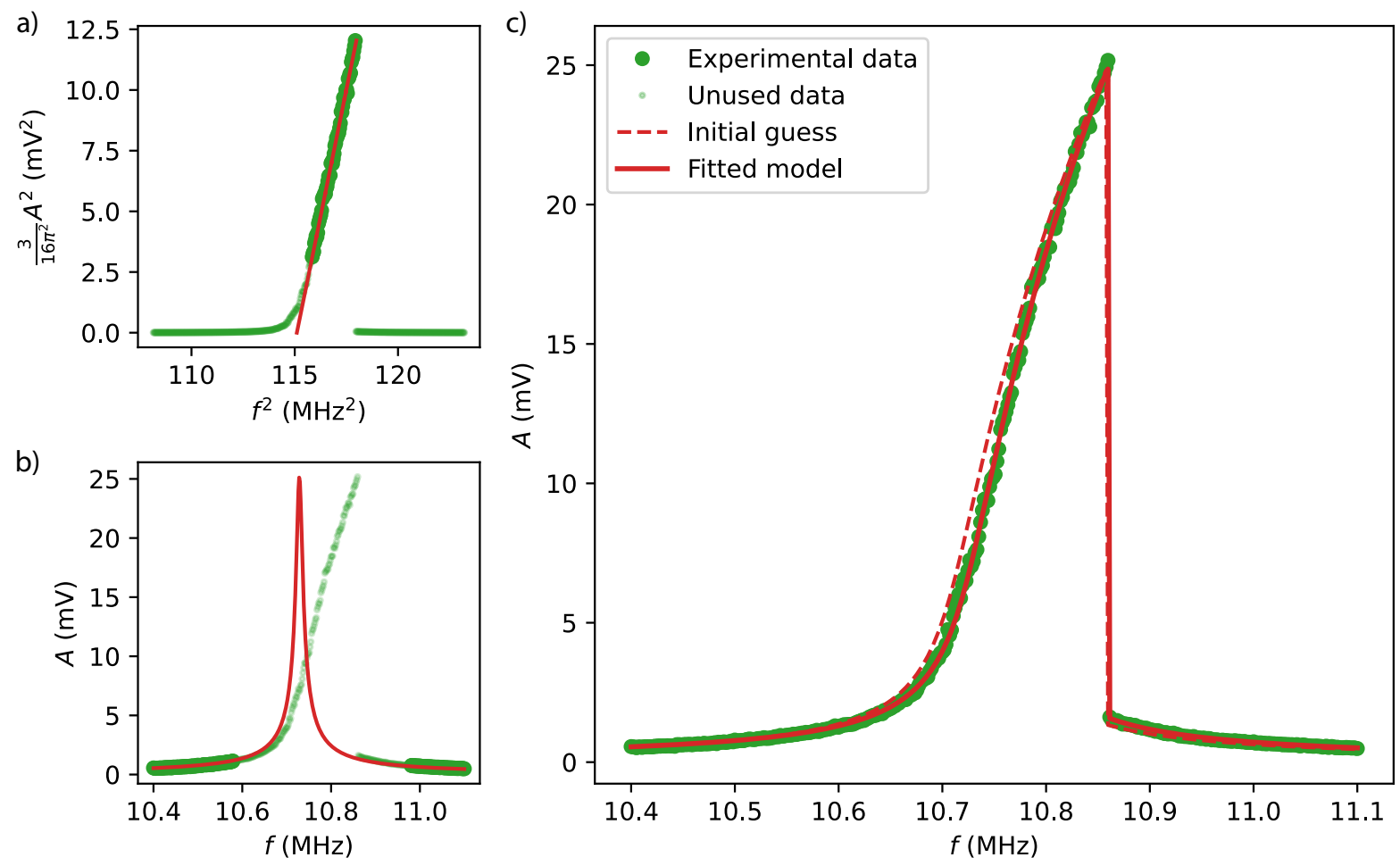


**Fig. 1** Fitting procedure applied on data seen in Fig. 1b. **a**, Fitting of backbone curve on high-amplitude data. **b**, Fitting of linear frequency response on off-resonance data. **c**, Final optimization of nonlinear frequency response.

$$w(r,t) = x(t)\phi(r), \tag{4}$$

with

$$\phi(r) = J_0\left(\alpha_0 \frac{r}{R}\right), \qquad \alpha_0 = 2.40483, \tag{5}$$

where $R$ is the membrane radius and $J_0$ denotes the Bessel function of the first kind. Next, we introduce a small static imperfection,

$$w_0(r) = \epsilon A \phi(r), \tag{6}$$

where $\epsilon \ll 1$ is a bookkeeping parameter and $A$ is the offset from flat configuration. The radial displacement of the membrane can be further written as

$$u(r,t) = u_0 r + \frac{r}{R}\left(1 - \frac{r}{R}\right) \sum_{k=1}^{N} q_k(t) \left(\frac{r}{R}\right)^{k-1}, \tag{7}$$

where the coefficients $q_k(t)$ are determined in terms of the generalized coordinate $x$ by minimizing the potential energy [1]. Moreover, the pre-strain is represented via

$$u_0 = \frac{n_0(1-\nu)}{Eh}, \tag{8}$$

with $n_0$ being the membrane pre-tension, $E$ the Young's modulus, $h$ the thickness, and $\nu$ the Poisson ratio. The nonlinear radial and circumferential strains assuming only axisymmetric vibrations are [1, 2]

$$\varepsilon_{rr} = \frac{\partial u}{\partial r} + \frac{1}{2}\left(\frac{\partial w}{\partial r}\right)^2 + \frac{\partial w}{\partial r}\frac{\partial w_0}{\partial r} - \alpha \Delta T(t), \tag{9}$$

$$\varepsilon_{\theta\theta} = \frac{u}{r} - \alpha \Delta T(t), \tag{10}$$

where $\alpha$ is the thermal expansion coefficient and $\Delta T(t)$ is the temperature difference, which can be obtained by solving the heat conduction equation

$$\frac{\partial \Delta T}{\partial t} + \frac{\Delta T}{\tau} = \frac{P_{\mathrm{F}}}{C_t}\cos(\omega_{\mathrm{F}} t) + \epsilon \frac{P_{\mathrm{S}}}{C_t}\cos(\omega_{\mathrm{S}} t), \tag{11}$$

where $C_t$ is the thermal capacitance, $\tau$ is the thermal time constant, and $P_{\mathrm{F}}$ and $P_{\mathrm{S}}$ denote the amplitudes of the fast and slow heating components, respectively. The Kirchhoff stresses in the membrane can be further obtained as

$$\sigma_{rr} = \frac{Eh}{1-\nu^2}(\varepsilon_{rr} + \nu\varepsilon_{\theta\theta}), \tag{12}$$

$$\sigma_{\theta\theta} = \frac{Eh}{1-\nu^2}(\varepsilon_{\theta\theta} + \nu\varepsilon_{rr}). \tag{13}$$

Knowing the displacement field and stresses in the membrane, one can obtain the kinetic ($T$) and potential energy ($U$) as follows:

$$\begin{aligned} T &= \frac{1}{2}\rho h \int_0^{2\pi}\int_0^R \left(\frac{\partial w}{\partial t}\right)^2 r\,dr\,d\theta, \\ U &= \frac{1}{2}\int_0^{2\pi}\int_0^R (\sigma_{rr}\varepsilon_{rr} + \sigma_{\theta\theta}\varepsilon_{\theta\theta})\, r\,dr\,d\theta, \end{aligned} \tag{14}$$

where $\rho$ is the density. Finally, substituting 14 in the Lagrange equations

$$\frac{d}{dt}\left(\frac{\partial T}{\partial \dot{x}}\right) - \frac{\partial T}{\partial x} + \frac{\partial U}{\partial x} = 0, \tag{15}$$

yields the following single-mode nonlinear equation of motion

$$\begin{aligned} &\epsilon^0\Big(4.9\, n_0\, x - \frac{5.8\, Eh\alpha\tau}{C_t} P_{\mathrm{F}}\cos(\omega_{\mathrm{F}} t)\, x + \frac{2.8\, Eh}{R^2}\, x^3 + 1.7\, R^2 h\rho\, \ddot{x}\Big) \\ &+\epsilon^1\Big(4.9\, An_0 - \frac{5.8\, AEh\alpha\tau}{C_t} P_{\mathrm{F}}\cos(\omega_{\mathrm{F}} t) - \frac{5.8\, Eh\alpha\tau}{C_t} P_{\mathrm{S}}\cos(\omega_{\mathrm{S}} t)\, x + \frac{8.5\, AEh}{R^2}\, x^2\Big) \\ &+\epsilon^2\Big(-\frac{5.8\, AEh\alpha\tau}{C_t} P_{\mathrm{S}}\cos(\omega_{\mathrm{S}} t) + \frac{5.7\, A^2 Eh}{R^2}\, x\Big) + O(\epsilon^3) = 0. \end{aligned} \tag{16}$$

Several observations follow directly from Eq. (16). First, the fast parametric term ($P_{\mathrm{F}}\cos(\omega_{\mathrm{F}} t)\, x$) appears as order $\epsilon^0$. Still, this term can be discarded in this work since the fast drive is placed close to resonance where principal parametric resonance condition is not met. Second, the fast direct ($P_{\mathrm{F}}\cos(\omega_{\mathrm{F}} t)$) and slow parametric ($P_{\mathrm{S}}\cos(\omega_{\mathrm{S}} t)\, x$) drives show up as order $\epsilon^1$. Finally, slow direct drive ($P_{\mathrm{S}}\cos(\omega_{\mathrm{S}} t)$) appears as order $\epsilon^2$, and therefore is negligible compared to the fast direct and slow parametric forcing.

This result justifies the modelling assumption for our experiment where we use only a fast direct drive and a slow parametric modulation, while neglecting other forcing terms. In the main text, we adopt a mass-normalized equation of motion for the fundamental mode, of the form

$$\ddot{x} + 2\Gamma\dot{x} + \left(\omega_0^2 + F_{\mathrm{S}}\cos(\omega_{\mathrm{S}}t)\right)x + \gamma x^3 = F_{\mathrm{F}}\cos(\omega_{\mathrm{F}}t). \tag{17}$$

Mass normalization is done by dividing Eq. (16) by the modal mass $m = 1.7R^2h\rho$. Furthermore, relaxation rate $\Gamma$ represents damping in the system, which is explicitly added to the equation to introduce dissipation. Linear resonance of the membrane is represented by $\omega_0^2$, which is predominantly dominated by pretension $n_0$ and slowly modulated at an amplitude $F_{\mathrm{S}}$ resulting from the temperature variation $\Delta T$. $\gamma$ denotes the cubic nonlinear stiffness, arising from geometric nonlinearity and scales as $Eh/R^2$. Finally, $F_{\mathrm{F}}$ is the amplitude of the fast direct force, applied as a result of the initial displacement $w_0$ and thus would be zero if the membrane was perfectly flat.

# 3 Derivation of the equation of motion in the rotating-frame

To derive the slow equations of motion (Eq. 2 of the main text), we start from Eq. (17) and apply the method of averaging. In the absence of the slow parametric drive ($F_{\mathrm{S}} = 0$), we assume a solution of the form $x(t) = P\cos(\omega_{\mathrm{F}}t) + Q\sin(\omega_{\mathrm{F}}t)$, together with the constraint $\dot{P}\cos(\omega_{\mathrm{F}}t) + \dot{Q}\sin(\omega_{\mathrm{F}}t) = 0$. Substituting this ansatz into Eq. (17) yields

$$\omega_F\left(-\dot{P}\sin\omega_{\mathrm{F}}t + \dot{Q}\cos\omega_{\mathrm{F}}t\right) = F(t) - (\omega_0^2 - \omega_{\mathrm{F}}^2)(P\cos(\omega_{\mathrm{F}}t) + Q\sin(\omega_{\mathrm{F}}t)) \tag{18}$$

$$- 2\Gamma\omega_{\mathrm{F}}\left(-P\sin\omega_{\mathrm{F}}t + Q\cos\omega_{\mathrm{F}}t\right) - \frac{3}{8}\frac{\gamma}{\omega_{\mathrm{F}}}(P\cos(\omega_{\mathrm{F}}t) + Q\sin(\omega_{\mathrm{F}}t))^3. \tag{19}$$

Defining the right hand side of this equation as $G(t)$, we can write this equation as a linear system for the quadrature velocities $\dot{P}$ and $\dot{Q}$:

$$\begin{cases} \dot{P} = -\dfrac{\sin(\omega_{\mathrm{F}}t)}{\omega_{\mathrm{F}}}G(t), \\ \dot{Q} = -\dfrac{\cos(\omega_{\mathrm{F}}t)}{\omega_{\mathrm{F}}}G(t). \end{cases} \tag{20}$$

Taking the average of these sets of equations gives the slow equations of motion in the absence of slow modulation of the resonance frequency:

$$\begin{cases} \dot{P} = \dfrac{\omega_0^2 - \omega_{\mathrm{F}}^2}{2\omega_{\mathrm{F}}}Q - \Gamma P + \dfrac{3}{8}\dfrac{\gamma}{\omega_{\mathrm{F}}}Q\left(P^2 + Q^2\right), \\ \dot{Q} = -\dfrac{\omega_0^2 - \omega_{\mathrm{F}}^2}{2\omega_{\mathrm{F}}}P - \Gamma Q - \dfrac{3}{8}\dfrac{\gamma}{\omega_{\mathrm{F}}}P\left(P^2 + Q^2\right) + \dfrac{F_{\mathrm{F}}}{2\omega_{\mathrm{F}}}. \end{cases} \tag{21}$$

To add back the slow modulation, we make the resonance frequency $\omega_0$ a time dependent variable, defined as $\hat{\omega}_0(t)$. Substitution of this term into Eq. (21) leads to Eq. 2 shown in the main text.

## 4 Libration frequency and parametric instability in slow dynamics

To obtain the linearized frequency corresponding to the stable fixed points N± in the slow dynamics, we start from the slow equation of motion (Eq. 2 of the main text) and perturb the fixed point ($P_0$, $Q_0$) as $P(t) = P_0 + p_0(t)$ and $Q(t) = Q_0 + q_0(t)$, and obtain the Jacobian

$$J(P_0, Q_0) = \begin{pmatrix} -\Gamma + \frac{3\gamma}{4\omega_\mathrm{F}} P_0 Q_0 & \frac{\hat{\omega}_0^2 - \omega_\mathrm{F}^2}{2\omega_\mathrm{F}} + \frac{3\gamma}{8\omega_\mathrm{F}} \left(P_0^2 + 3Q_0^2\right) \\ -\frac{\hat{\omega}_0^2 - \omega_\mathrm{F}^2}{2\omega_\mathrm{F}} - \frac{3\gamma}{8\omega_\mathrm{F}} \left(3P_0^2 + Q_0^2\right) & -\Gamma - \frac{3\gamma}{4\omega_\mathrm{F}} P_0 Q_0 \end{pmatrix}. \tag{22}$$

Evaluating it for the unmodulated resonator ($\hat{\omega}_0 = \omega_0$), we determine the trace and determinant as

$$\mathrm{tr}(J) = -2\Gamma, \tag{23}$$

$$\det(J) = \Gamma^2 + \left(\frac{\omega_0^2 - \omega_\mathrm{F}^2}{2\omega_\mathrm{F}} + \frac{3}{8}\frac{\gamma}{\omega_\mathrm{F}} R_0^2\right)\left(\frac{\omega_0^2 - \omega_\mathrm{F}^2}{2\omega_\mathrm{F}} + \frac{9}{8}\frac{\gamma}{\omega_\mathrm{F}} R_0^2\right). \tag{24}$$

where $R_0 = \sqrt{P_0^2 + Q_0^2}$. The eigenvalues $\lambda_i$ are therefore

$$\begin{aligned} \lambda_i &= \frac{\mathrm{tr}(J) \pm \sqrt{\mathrm{tr}(J)^2 - 4\det(J)}}{2} \\ &= -\Gamma \pm i\sqrt{\left(\frac{\omega_0^2 - \omega_\mathrm{F}^2}{2\omega_\mathrm{F}} + \frac{3}{8}\frac{\gamma}{\omega_\mathrm{F}} R_0^2\right)\left(\frac{\omega_0^2 - \omega_\mathrm{F}^2}{2\omega_\mathrm{F}} + \frac{9}{8}\frac{\gamma}{\omega_\mathrm{F}} R_0^2\right)}, \end{aligned} \tag{25}$$

The imaginary part $\omega_\mathrm{L} = \mathrm{Im}(\lambda_i)$ defines the libration frequency associated with small oscillations around a stable slow-flow fixed point. For the drum studied in the main text, $f_\mathrm{L} = \omega_\mathrm{L}/(2\pi)$ is 31.2 kHz for the high-amplitude stable fixed point $N+$ and 36.5 kHz for the low-amplitude stable fixed point $N-$. Since the modulation is initiated from $N+$, we consider the corresponding libration frequency $f_\mathrm{L} = 31.2$ kHz in the following.

A more minimalistic analysis of the slow equations of motion can be performed to illustrate the effects of the slow modulation on the slow dynamics. This is done by discarding all higher-order terms in $P$ and $Q$, solving the equation for $\dot{P}$ in the absence of modulation for $Q$ and substituting the result into the equation for $\dot{Q}$. This gives a second order differential equation of the form

$$\ddot{P} + 2\Gamma\dot{P} + \left(\Gamma^2 + \left(\frac{\omega_0^2 - \omega_\mathrm{F}^2}{2\omega_\mathrm{F}}\right)^2\right) P = \left(\frac{\omega_0^2 - \omega_\mathrm{F}^2}{2\omega_\mathrm{F}}\right)\frac{F_\mathrm{F}}{2\omega_\mathrm{F}}. \tag{26}$$

We then introduce slow modulation by replacing $\omega_0^2$ with $\omega_0^2 + F_\mathrm{S}\cos(\omega_\mathrm{S} t)$. As a result, the effective slow stiffness becomes

$$\left[\Gamma^2 + \left(\frac{\omega_0^2 - \omega_\mathrm{F}^2}{2\omega_\mathrm{F}}\right)^2 + \frac{F_\mathrm{S}^2}{8\omega_\mathrm{F}^2}\right] + \left(\frac{F_\mathrm{S}\omega_0^2}{2\omega_\mathrm{F}^2} - \frac{F_\mathrm{S}}{2}\right)\cos(\omega_\mathrm{S} t) + \frac{F_\mathrm{S}^2}{8\omega_\mathrm{F}^2}\cos(2\omega_\mathrm{S} t) \tag{27}$$

and the right hand side of the equation acquires a harmonic term,

$$\left(\frac{\omega_0^2 - \omega_\mathrm{F}^2}{2\omega_\mathrm{F}}\right)\frac{F_\mathrm{F}}{2\omega_\mathrm{F}} + \frac{F_\mathrm{F} F_\mathrm{S}}{4\omega_\mathrm{F}^2}\cos(\omega_\mathrm{S} t). \tag{28}$$

We note that this term originates from the slow parametric modulation, and is therefore distinct from the slow direct drive discussed in Supplementary Note 2.

The slow dynamics are thus parametrically driven at $\omega_\mathrm{S}$ and $2\omega_\mathrm{S}$, which is of similar structure to the Mathieu equation. Assuming this structure carries over to the nonlinear slow dynamics, we expect the slow dynamics to activate parametric instability tongues when $\omega_\mathrm{S} \approx \frac{2\omega_\mathrm{L}}{n}$ with $n \in \mathbb{N}$. For the high-amplitude fixed point N+, from which all experiments are started, the first six tongues are expected at $f_\mathrm{S} \approx$ 62.4 kHz, 31.2 kHz, 20.8 kHz, 15.6 kHz, 12.5 kHz and 10.4 kHz. In addition, Eq. (28) predicts a direct resonance of the slow dynamics at $\omega_\mathrm{L}$.

Fig. 2 shows the observed dynamics when the slow dynamics are parametrically driven at the tongue frequencies for $n = 2, \ldots, 5$. Simulations are performed at a constant $A_\mathrm{S} = 40$ kHz, measurements are matched manually from data shown in Fig. 3. For all responses, Poincaré points (black dots) are shown by sampling the data at time intervals equal to the modulation period, starting from the time at the largest observed amplitude. Period doubled or chaotic motion is observed following the Poincaré points for each response, which is consistent with the expectation of instability tongues at these frequencies. Interestingly, the primary parametric instability ($n = 1$) was not activated in simulation and the corresponding frequency could not be measured due to constraints in acquisition rate.

To investigate the behaviour at $f_\mathrm{S} = 2f_\mathrm{L}/n$ further, we numerically sweep the modulation amplitude $A_\mathrm{S}$ for $n = 1, \ldots, 6$ (Fig. 4). Grey lines indicate the range of the response amplitude. Again, only the higher-order instabilities ($n > 1$) are activated. For each $n > 1$ an onset modulation amplitude is observed, which increases with $n$ – thus a larger slow force is required to activate higher order parametric instabilities (note that this does not hold for $V_\mathrm{S}$ due to the nonlinear relation with $A_\mathrm{S}$ and $F_\mathrm{S}$). Moreover, the range of $A_\mathrm{S}$ in which period doubling and chaos appear shrinks with increasing $n$, suggesting that lower-order instabilities are more easily activated. The primary instability ($n = 1$) remains the exception. A possible reason is that $2f_\mathrm{L} =$ 62.4 kHz lies too far above the relaxation rate $f_\mathrm{relax}$, and therefore outside the critical regime $f_\mathrm{S} \approx f_\mathrm{relax}$ in which the instabilities are observed.

Finally, we numerically sweep the modulation frequency $f_\mathrm{S}$ at several fixed, arbitrarily chosen amplitudes $A_\mathrm{S}$ (Fig. 5). The Poincaré maps, together with lines marking the tongue frequencies, show that period-doubled and chaotic motion appear close

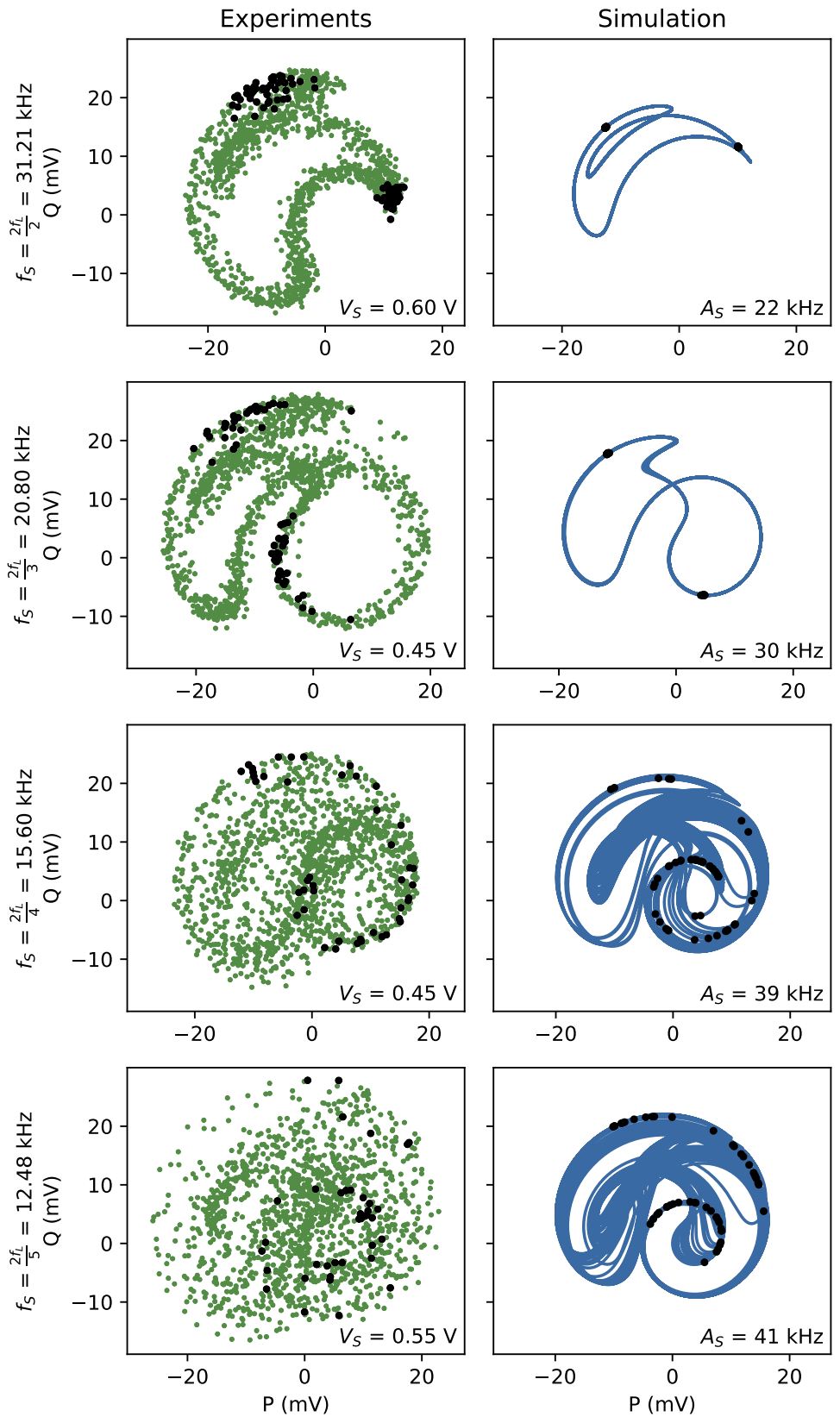


**Fig. 2** Experimental and numerical response at parametric instability frequencies for $n = 2, 3, 4, 5$. Simulations are modulated at $A_{\mathrm{S}}$ slightly above the instability threshold, experiments have been chosen from Fig. 3 to visually match the simulated response. Slow drive amplitudes are shown in the panels. Black dots show the Poincaré points taken at time intervals equal to the modulation period.

to $f_{\mathrm{S}} = 2f_{\mathrm{L}}/n$ for $n > 1$. At the lowest amplitude, $A_{\mathrm{S}} = 1\,\mathrm{kHz}$, a directly driven resonance peak is visible, consistent with Eq. (28).

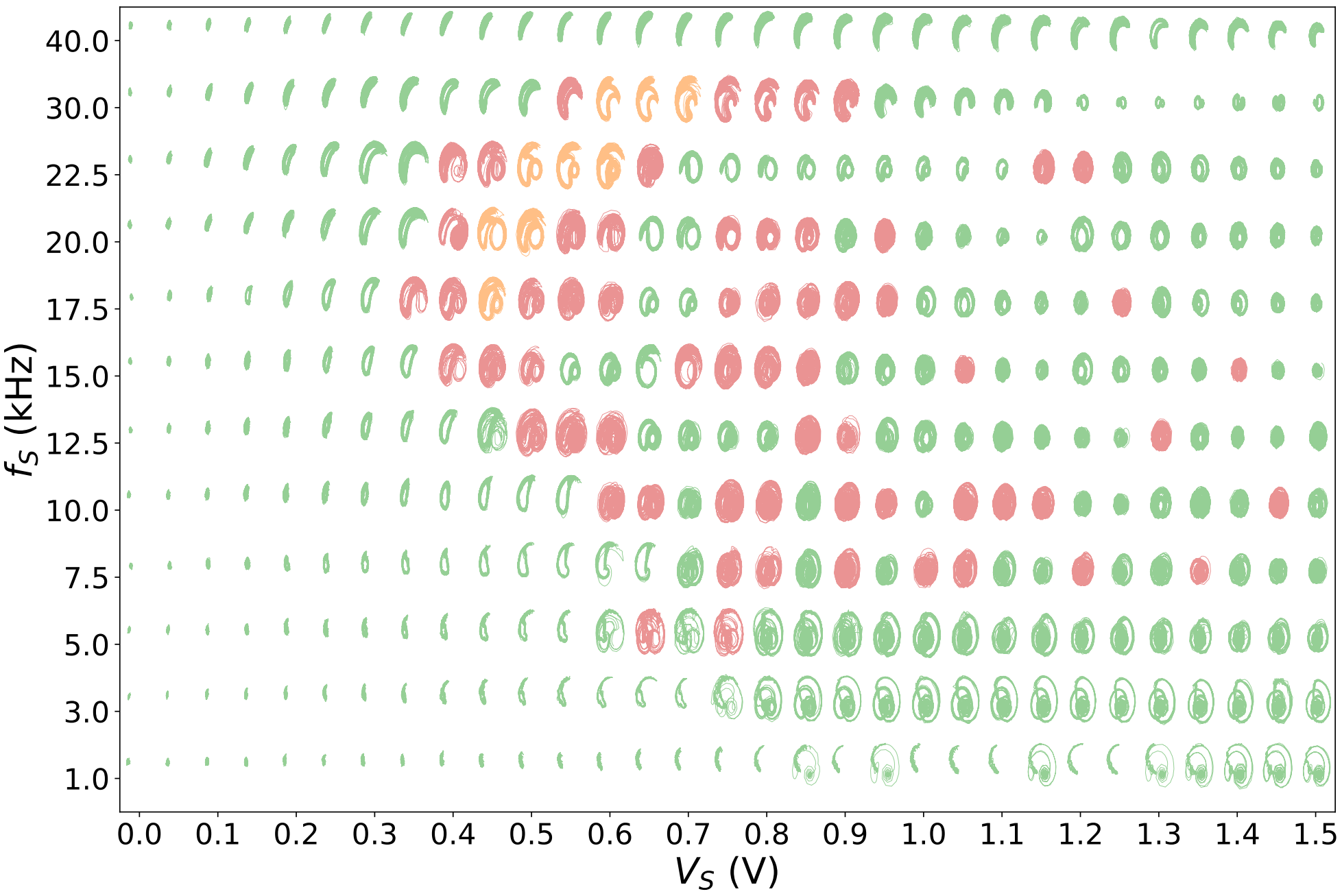


**Fig. 3** Experimental parameter sweep on the drum used in the main text. The slow modulation amplitude $V_{\mathrm{S}}$ is swept up from 0 V to 1.5 V, which is repeated for different slow modulation frequencies $f_{\mathrm{S}}$. Colors indicate the type of observed dynamics, where green indicates periodic motion, yellow period-doubled motion and red chaotic motion.

# 5 Lyapunov exponent calculation

To calculate the Largest Lyapunov exponent from experimental data, we apply the Rosenstein algorithm [3]. This method is known to work well with short and noisy datasets, since it determines the average separation of points that are initially close together. This is done by first reconstructing the attractor dynamics using the method of delays, which introduces the embedding dimension $m$ and reconstruction delay $L$. As described by Rosenstein [3], generally these parameters should be chosen such that $m > 2n$ where $n$ is the number of states in the system and $L$ should be approximately where the autocorrelation function drops below $1 - \frac{1}{e}$.

After reconstructing the attractor dynamics, the nearest neighbour for each point of the reconstructed dynamics is computed, and their separation $s_j(i)$, where $j$ indicates the starting point and $i$ indicates the discrete time steps, is calculated by taking the Euclidean norm as the points evolve in time. By taking the average of the natural logarithm of each separation function $\langle \ln s_j(i) \rangle$, the effect of stochastic variations are then minimized. The separation function, when the dynamics are chaotic, is expected to show an exponential increase, followed by a plateau as the dynamics are bounded by the attractor. The initial increase in separation is determined by the largest Lyapunov exponent $\lambda_{\max}$ as

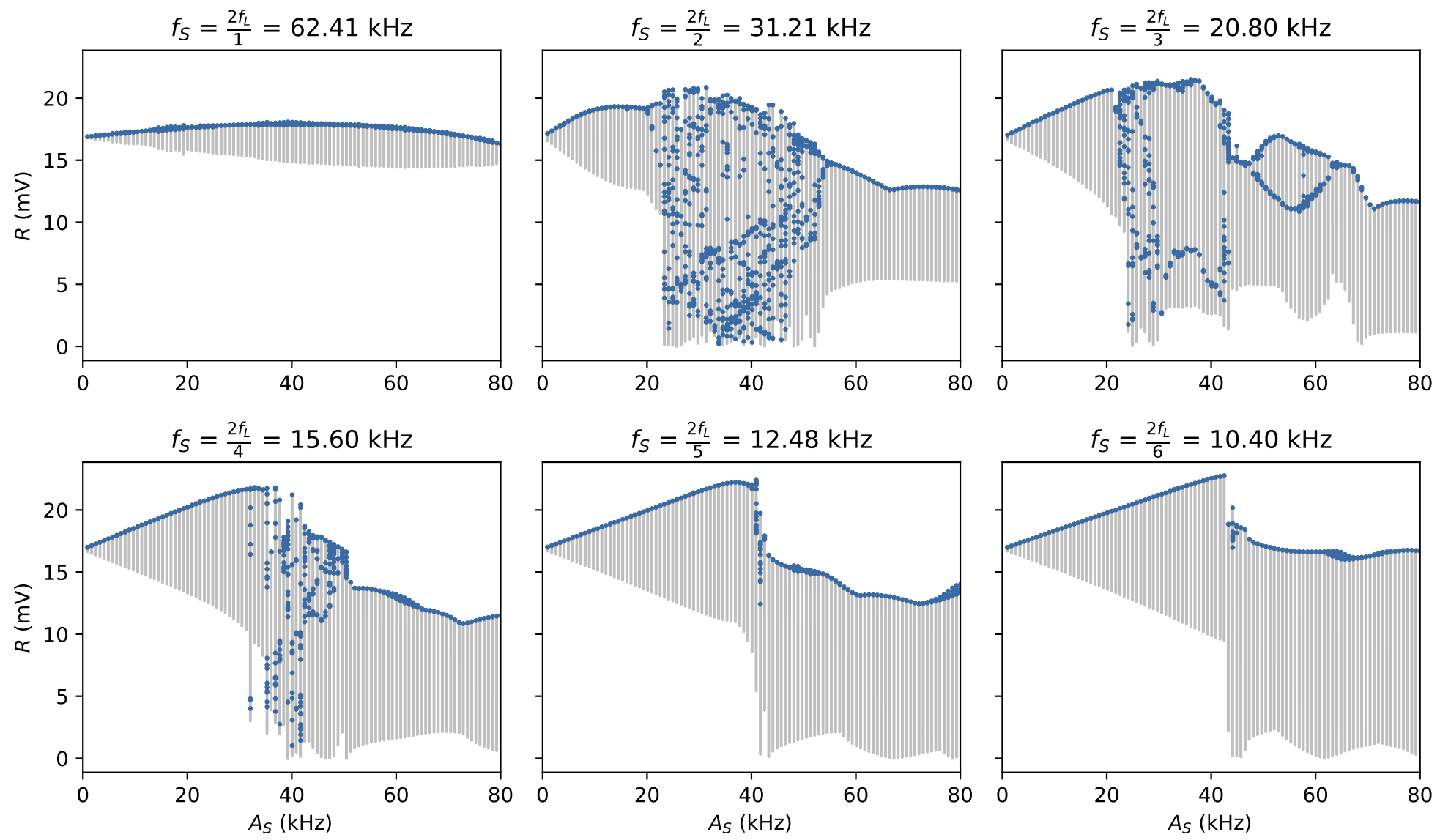


**Fig. 4** Simulated Poincaré maps of slow modulation amplitude ($A_{\mathrm{S}}$) sweeps at different modulation frequencies ($f_{\mathrm{S}}$) when these satisfy parametric instability conditions. Poincaré points (blue dots) have been taken with respect to the maximum amplitude, at a time interval equal to the slow modulation period. Grey lines indicate the amplitude range at the modulation amplitude.

$$s_j(i) = C_j e^{\lambda_{\max}(i\Delta t)}, \tag{29}$$

where $C_j$ is the initial separation and $\Delta t$ is the time step of the experimental data. By taking the natural logarithm, Eq. (29) can be rewritten as

$$\langle \ln s_j(i) \rangle = \langle \ln C_j \rangle + \lambda_{\max}(i\Delta t) \tag{30}$$

where $\langle \cdot \rangle$ denotes the average. From this equation, $\lambda_{\max}$ can be obtained by performing a linear fit on the experimental average separation. Here, this is done in Python, using numpy's `polyfit()` function with degree 1.

The algorithm is applied to data shown in Fig. 2, 3 and 4 in the main text, to compare the average separation plots. Here, we use $m = 10 > 2n = 4$ and $L = 6$, which is slightly higher than the value suggested by the approximation using the autocorrelation function ($L = 3$). The results are shown in Fig. 6. While all curves show an increase at first, the slope of the chaotic data is much larger. We note that due to noise in our experiments, a decrease in separation is highly unlikely to occur even for a periodic response.

To show the effect of changes in $L$ and $m$, different separation plots are shown in Fig. 7. The plot resulting from the same values as before is highlighted in green. Fig. 7 shows the chosen $L$ and $m$ values are chosen such that varying these hyper parameters do not greatly influence the shape of the average separation plots, which suggests these values are close to optimal.

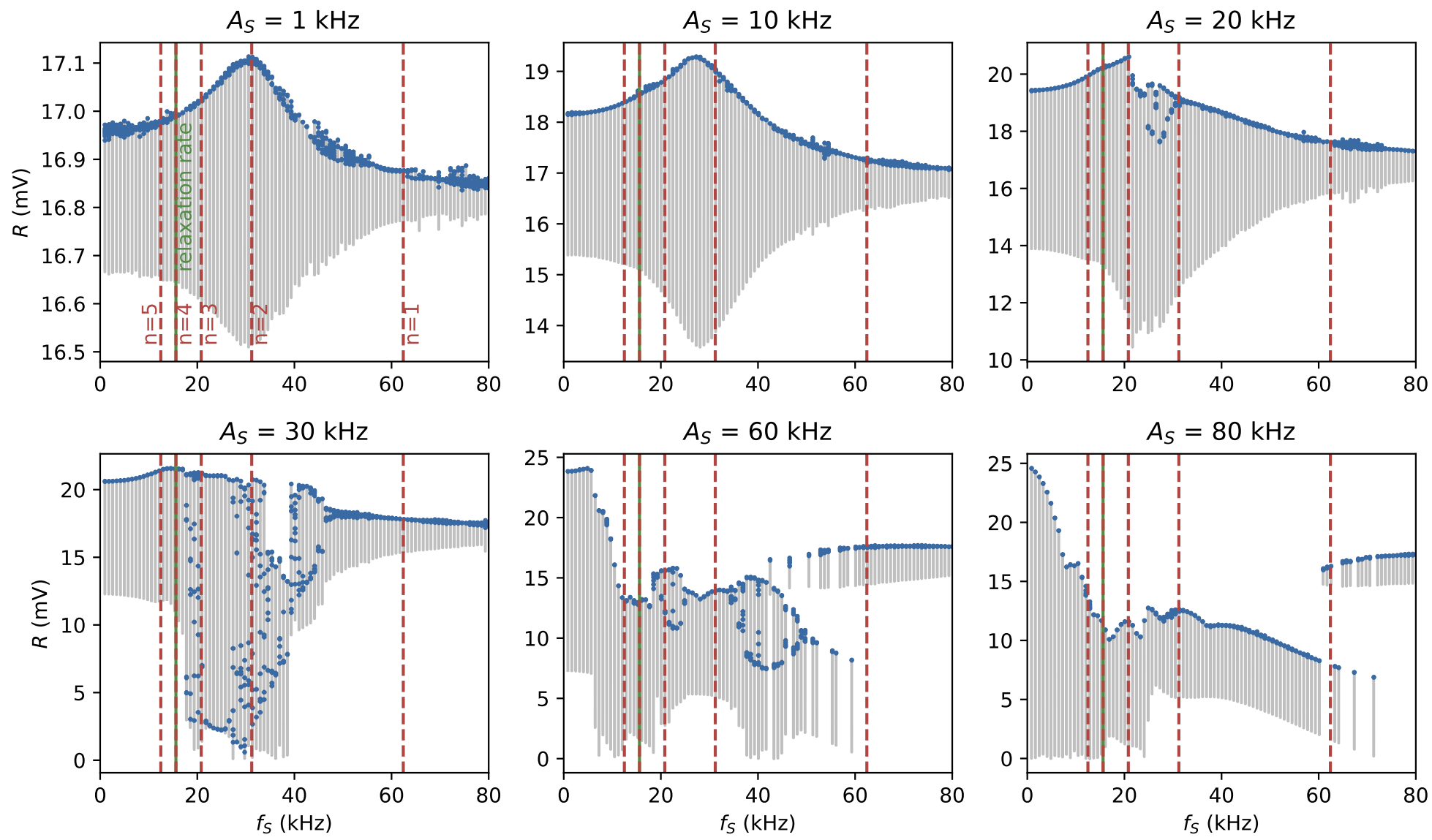

**Fig. 5** Simulated Poincaré maps of slow modulation frequency ($f_S$) sweeps at different modulation amplitudes ($A_S$). Poincaré points (blue dots) have been taken with respect to the maximum amplitude, at a time interval equal to the slow modulation period. Grey lines indicate the amplitude range at the modulation frequency.

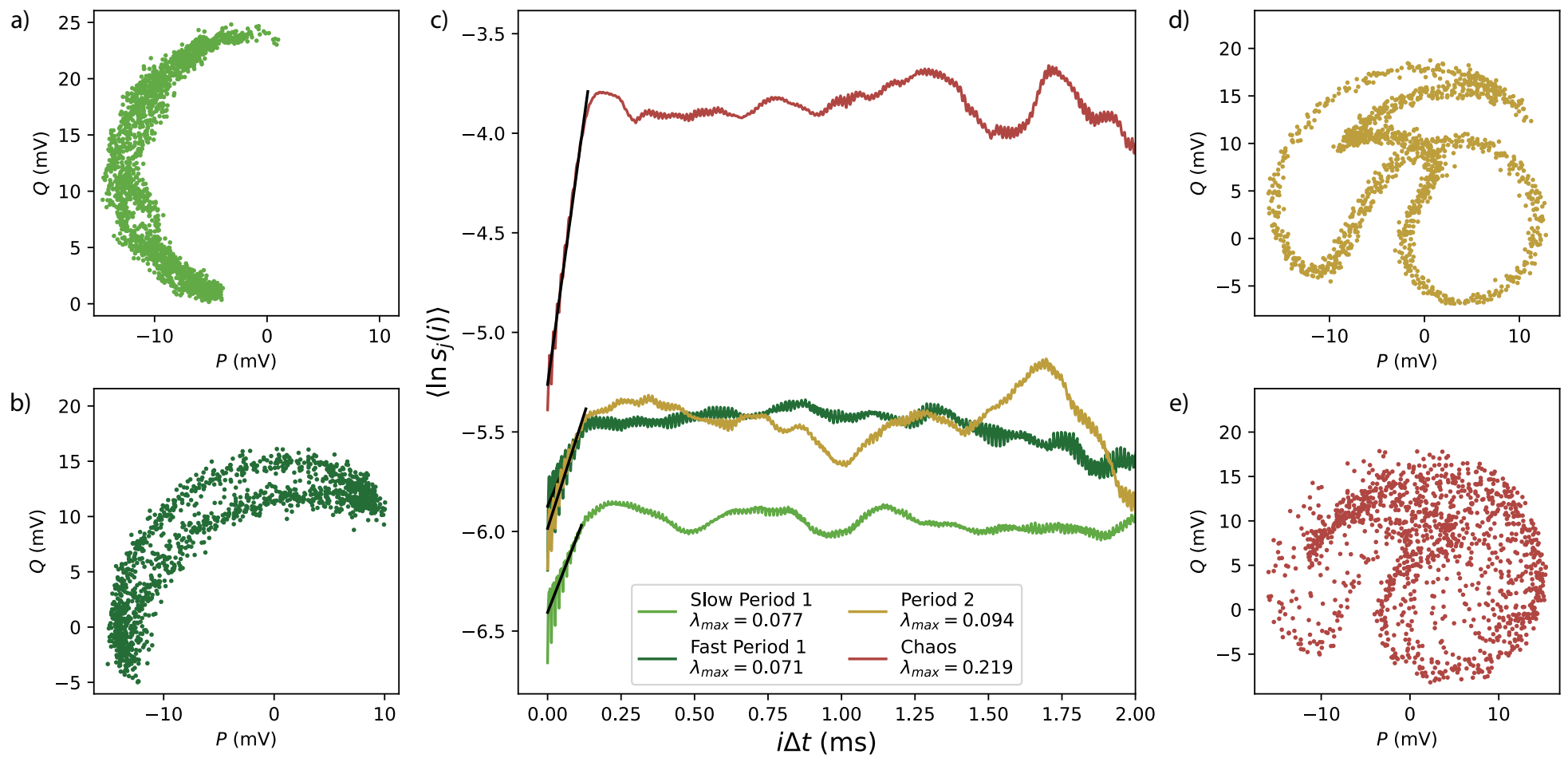

**Fig. 6** Results from the Rosenstein analysis on experimental results shown in the main text. The middle panel shows the resulting separation plots for the different measurement data sets, calculated using $L = 6$ and $m = 10$. The slow drive parameters used for each measurement are a) $V_S = 1\,\mathrm{V}$, $f_S = 1\,\mathrm{kHz}$, b) $V_S = 1\,\mathrm{V}$, $f_S = 50\,\mathrm{kHz}$, d) $V_S = 0.28\,\mathrm{V}$, $f_S = 20\,\mathrm{kHz}$, e) $V_S = 0.32\,\mathrm{V}$, $f_S = 20\,\mathrm{kHz}$,

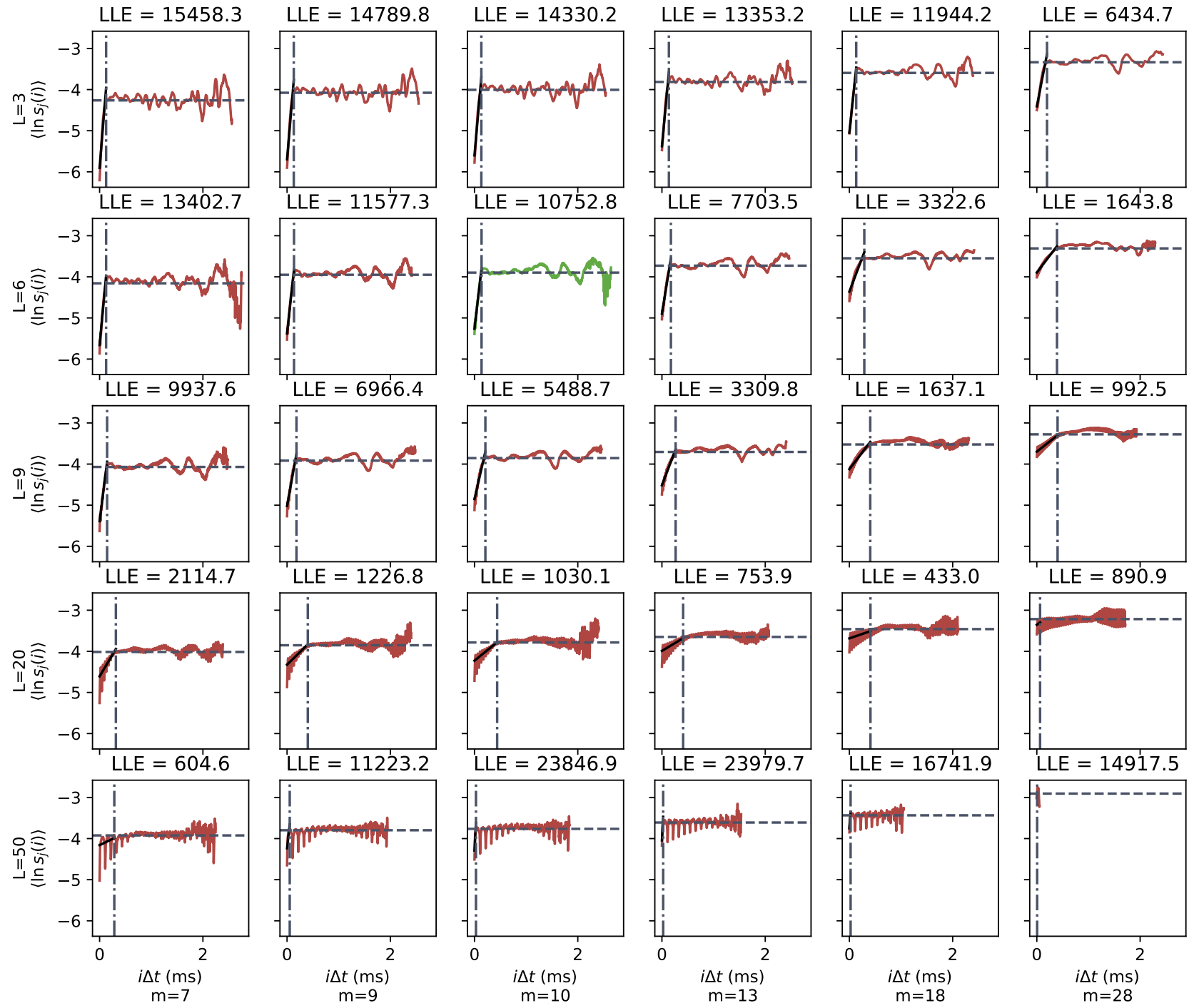


**Fig. 7** Effects of changes in $m$ and $L$ to the resulting average separation plots. The plot highlighted in green corresponds to the settings used for the final largest Lyapunov extraction.

As additional proof, we perform the same analysis on numerical data that represent our experiments. Again, the data shown in Fig. 2, 3 and 4 of the main text is used, and the same $m = 10$ and $L = 6$ are set in the Rosenstein algorithm. The results are shown in Fig. 8, where only for the chaotic dataset a positive $\lambda_{\max}$ is found. We note the substantial difference in Lyapunov exponents from experimental and numerical data, which is likely caused by noise in the measured dynamics or slight differences in parameters. Still, the qualitative similarity between experimental and numerical PQ-spaces supports the conclusion of chaotic dynamics for the data shown in panel e in both Fig. 6 and Fig. 8.

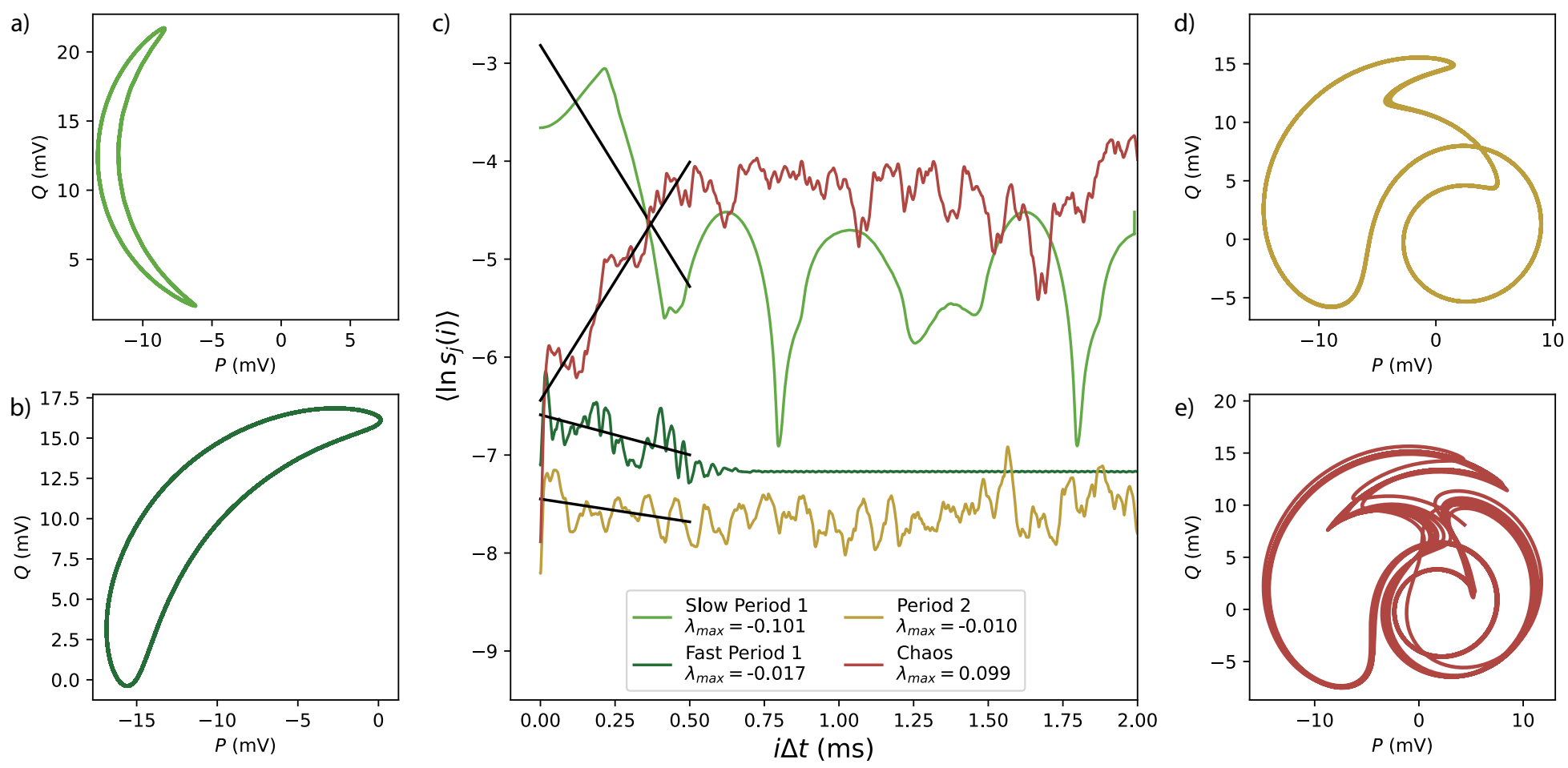


**Fig. 8** Results from the Rosenstein analysis on the numerical results shown in the main text. The middle panel shows the resulting separation plots for the different numerical data sets, calculated using $L = 6$ and $m = 10$. The slow drive parameters used for each measurement are a) $A_S = 60\,\text{kHz}$, $f_S = 1\,\text{kHz}$, b) $A_S = 30\,\text{kHz}$, $f_S = 50\,\text{kHz}$, d) $A_S = 27.16\,\text{kHz}$, $f_S = 20\,\text{kHz}$, e) $A_S = 31.04\,\text{kHz}$, $f_S = 20\,\text{kHz}$,

# 6 Bifurcation diagram and frequency spectra

To further support the identification of period-doubled and chaotic oscillations, we construct bifurcation diagrams from Poincaré sections of the measured and simulated $(P, Q)$ trajectories. To construct the Poincaré map, we sample the amplitude once per modulation cycle: starting from the point of maximum amplitude in the time series, we step forward and backward in time in increments of the modulation period $\frac{1}{f_S}$, recording the amplitude at each step. For the measurements in this section, a measurement time of 3 ms result in 60 Poincaré points. The same simulation time has been set for numerical results, to more easily compare the results.

Figure 9a shows the experimental bifurcation diagram corresponding to the data presented in Fig. 5a and b of the main text, while Fig. 9j shows the corresponding numerical result.

We also obtain the frequency spectra of the main text measurements, shown in Fig. 2, 3 and 4, by applying the fast Fourier transform to the acquired temporal data. These are shown in Fig. 10. Here, it can be seen that the slow modulation generates a frequency comb around the fast drive tone placed at $f_F$, with spacings between the teeth equal to the slow modulation frequency $f_S$ in Fig. 10a and b. When period doubling occurs the spacing is halved, as seen in Fig. 10c, and for chaos the frequency content appear random like as shown in Fig. 10d.

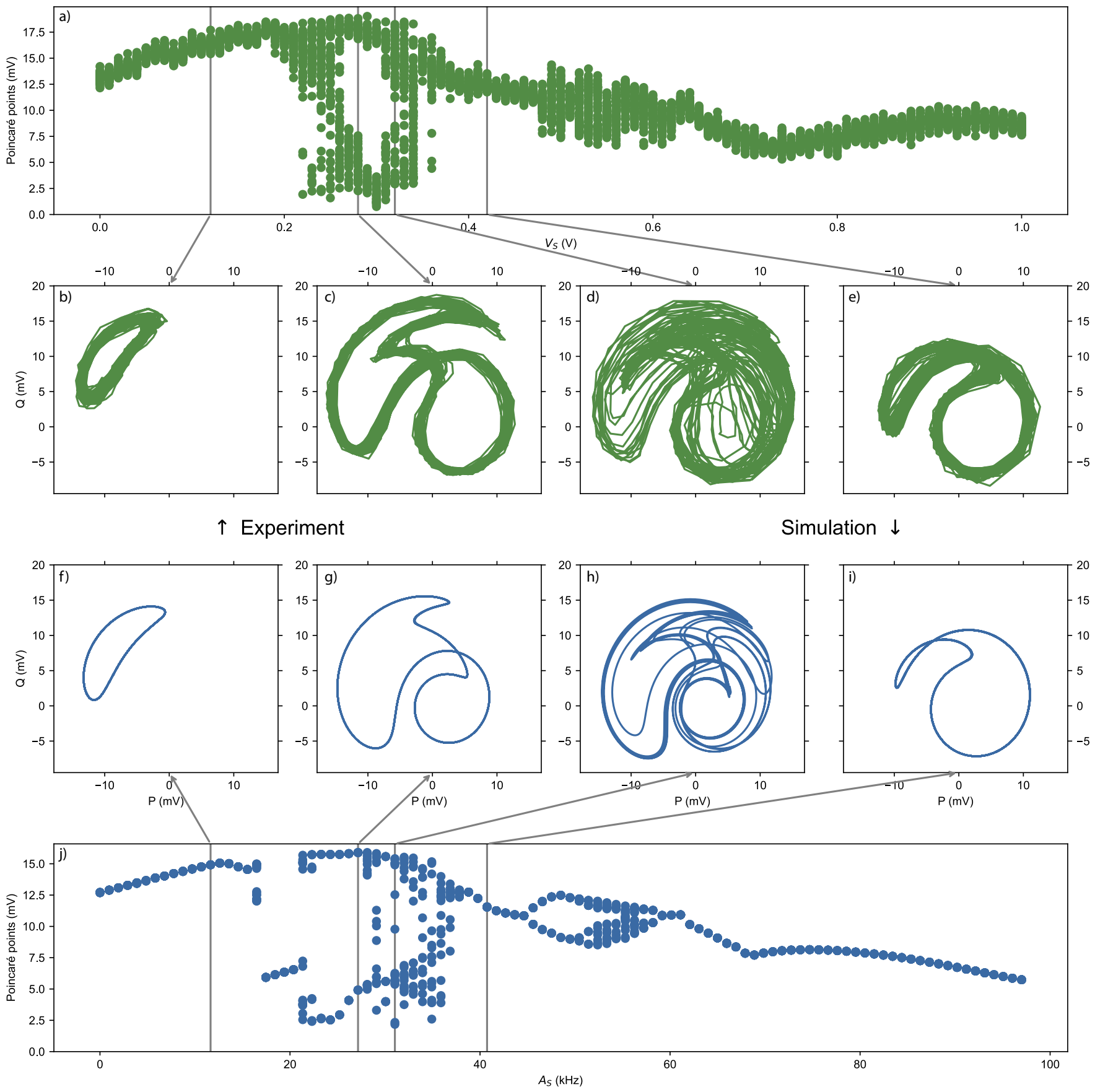


**Fig. 9** Bifurcation diagram for the drum in the main text. **a, j**, Experimental and simulated bifurcation diagrams. **b-i**, highlighted P-Q spaces at specific drive levels.

# 7 Melnikov's method

A simple analytical tool for predicting the onset of chaos in weakly perturbed nonlinear systems is based on the Melnikov method. Here, we apply this approach to identify the parameter threshold at which the stable and unstable manifolds of the saddle intersect transversely.

To this end, we introduce the action-angle variables

$$J = \frac{P^2 + Q^2}{2}, \qquad \psi = \tan^{-1}(Q/P), \tag{31}$$

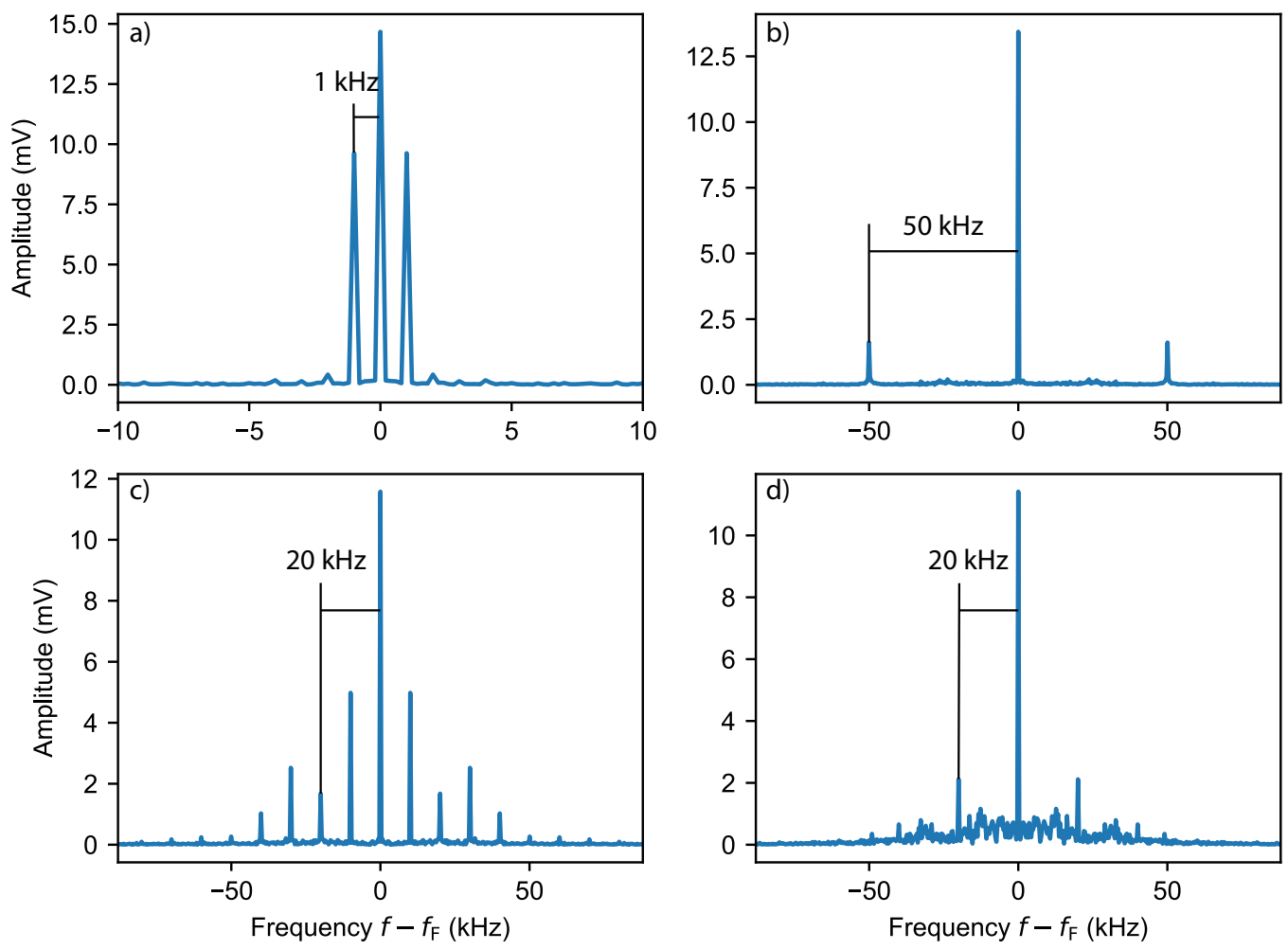


**Fig. 10** Fast Fourier transform on data shown in the main paper. **a**, Slow parametric modulation at 1 kHz, leading to adiabatic tracking. **b**, Fast parametric modulation at 50 kHz, leading to delay in tracking. **c**, Period doubled dynamics, at $f_S = 20$ kHz. **d**, Chaotic dynamics, at $f_S = 20$ kHz.

which transform the physical quadrature system (21) into the canonical form

$$\begin{cases} \dot{J} = -F_F\sqrt{2J}\sin\psi - 2\Gamma_1\omega_F J, \\ \dot{\psi} = -\Omega + \dfrac{3}{2}\gamma J - \dfrac{F_F\sqrt{2J}}{\cos\psi} + F_S\cos(\nu_0 t). \end{cases} \tag{32}$$

where $\Gamma_1 = 2\Gamma$, $\nu_0 = 2\omega_F\omega_S$, $\Omega = (\omega_F^2 - \omega_0^2)$. The system in Eq. (32) can be written in the standard form

$$\dot{\mathbf{x}} = \mathbf{f}(\mathbf{x}) + \varepsilon\mathbf{g}(\mathbf{x}, t), \qquad \mathbf{x} = \begin{pmatrix} J \\ \psi \end{pmatrix} \tag{33}$$

with

$$\mathbf{f}(\mathbf{x}) = \begin{pmatrix} -F_F\sqrt{2J}\sin\psi \\ -\Omega + \dfrac{3}{2}\gamma J - \dfrac{F_F\sqrt{2J}}{\cos\psi} \end{pmatrix}, \qquad \mathbf{g}(\mathbf{x}, t) = \begin{pmatrix} -2\Gamma_1\omega_F J \\ F_S\cos(\nu_0 t) \end{pmatrix}. \tag{34}$$

Here, $\varepsilon$ is a bookkeeping parameter introduced to characterize the weak perturbation due to damping and slow harmonic modulation, while $\dot{\mathbf{x}} = \mathbf{f}(\mathbf{x})$ denotes the corresponding unperturbed Hamiltonian system, with Hamiltonian energy

$$H(J, \psi) = -\Omega J + \frac{3}{4}\gamma J^2 - F_F\sqrt{2J}\cos\psi. \tag{35}$$

The latter possesses a hyperbolic saddle equilibrium with stable and unstable manifolds forming the homoclinic separatrix loops to be analysed. Let $q_0(t)$ denote a homoclinic orbit of the unperturbed system connecting the hyperbolic saddle equilibrium $p_0(t)$ to itself. The Melnikov function associated with this orbit is defined as

[4]

$$M(t_0) = \int_{-\infty}^{\infty} \mathbf{f}(q_0(t)) \wedge \mathbf{g}(q_0(t), t + t_0)\, dt, \tag{36}$$

The parameter $t_0$ accounts for the phase shift of the periodic perturbation relative to the homoclinic orbit. For sufficiently small $\epsilon$, the Melnikov function provides the leading-order approximation of the distance between the perturbed stable and unstable manifolds,

$$d(t_0) = \frac{\varepsilon M(t_0)}{|\mathbf{f}(q_0(0))|} + O(\varepsilon^2). \tag{37}$$

Hence, a simple zero of $M(t_0)$ implies a transverse intersection of the stable and unstable manifolds. Such a transverse homoclinic intersection corresponds to the destruction of the separatrix barrier and signals the onset of chaotic dynamics [5].

In the present problem, the unperturbed system possesses two distinct homoclinic orbits. The corresponding Melnikov functions are therefore obtained by evaluating Eq. (36) along the two homoclinic branches,

$$\begin{aligned} M_{\pm}(t_0) = \int_{-\infty}^{\infty} \Bigg\{ & \left[-F_{\mathrm{F}} \sqrt{2J_{\pm}(t)} \sin\psi_{\pm}(t)\right] \cdot F_{\mathrm{S}} \cos(\nu_0(t + t_0)) \\ & - \left[-\Omega + \frac{3}{2}\alpha J_{\pm}^2(t) - \frac{\gamma\cos\phi_{\pm}(t)}{\sqrt{2J_{\pm}(t)}}\right] [-2\Gamma_1 \omega_{\mathrm{F}} J_{\pm}] \Bigg\}\, dt. \end{aligned} \tag{38}$$

Here, the subscripts $+$ and $-$ refer to the outer and inner homoclinic branches, respectively. The integrals in Eq. (38) can be evaluated analytically using the residue method [6], yielding

$$M_{+}(t_0) = -\frac{8\pi F_{\mathrm{S}}\nu_0}{3\gamma} \frac{\sinh\left(\frac{4\nu_0\theta_0}{3\rho\gamma}\right)}{\sinh\left(\frac{4\nu_0}{3\rho\gamma}\right)} \sin(\nu_0 t_0) - \frac{16\Gamma_1\omega_{\mathrm{F}}}{3\gamma}\left(\Omega\theta_0 - \frac{9}{8}\rho\gamma\right), \tag{39}$$

and

$$M_{-}(t_0) = -\frac{8\pi F_{\mathrm{S}}\nu_0}{3\gamma} \frac{\sinh\left(\frac{4\nu_0\theta_0}{3\rho\gamma}\right)}{\sinh\left(\frac{4\nu_0}{3\rho\gamma}\right)} \sin(\nu_0 t_0) + \frac{16\Gamma_1\omega_{\mathrm{F}}}{3\gamma}\left(\Omega(\pi - \theta_0) + \frac{9}{8}\rho\gamma\right). \tag{40}$$

Here, $\theta_0 = \arccos[(r_+ + r_-)/(r_+ - r_-)]$ and $r_{\pm} = 2\left(\left(\frac{2\Omega}{3\gamma} - j_2\right) \pm \sqrt{2j_2\left(\frac{2\Omega}{3\gamma} - j_2\right)}\right)$ where $(J, \psi) = (j_2, \pi)$ denotes the hyperbolic saddle equilibrium. The existence of zeros of $M_{\pm}(t_0)$ is determined by the competition between the oscillatory contribution associated with the slow modulation and the constant contribution due to damping.

We set $M_{\pm}(t_0) = 0$ and introduce the auxiliary threshold functions

$$R_{+} = \frac{|8\Omega\theta_0 - 9\rho\gamma|}{4\pi\nu_0} \frac{\sinh\left(\dfrac{4\pi\nu_0}{3\rho\gamma}\right)}{\sinh\left(\dfrac{4\psi_0\nu_0}{3\rho\gamma}\right)} \Gamma_1\omega_{\mathrm{F}}, \tag{41}$$

and

$$R_{-} = \frac{|8\Omega(\pi - \theta_0) + 9\rho\gamma|}{4\pi\nu_0} \frac{\sinh\left(\dfrac{4\pi\nu_0}{3\rho\gamma}\right)}{\sinh\left(\dfrac{4(\pi - \theta_0)\nu_0}{3\rho\gamma}\right)} \Gamma_1\omega_{\mathrm{F}}. \tag{42}$$

Accordingly, transverse intersection of the stable and unstable manifolds occurs when

$$F_{\mathrm{S}} > T_{\pm}. \tag{43}$$

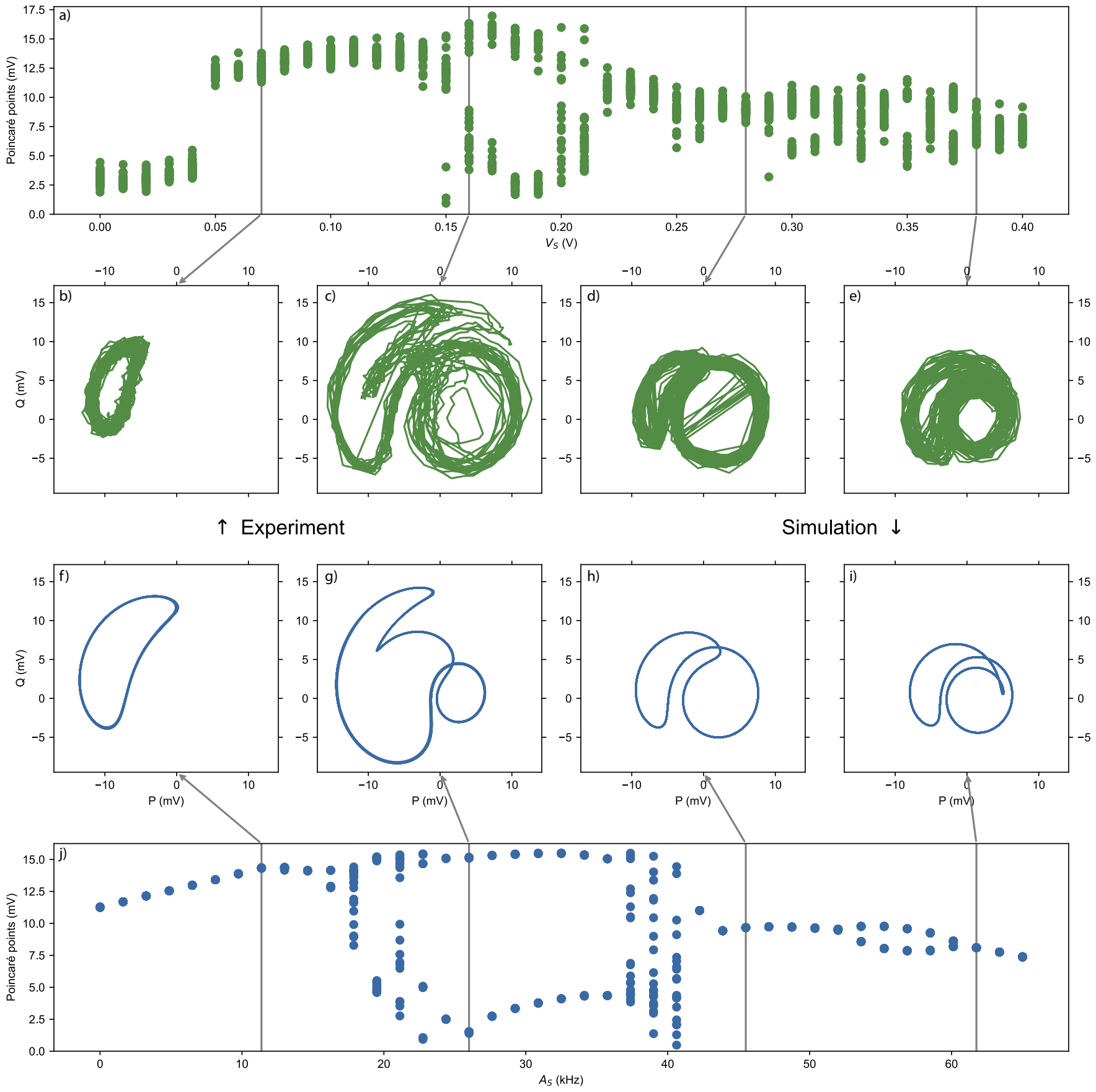


**Fig. 11** Bifurcation diagram for an additional drum to the one in the main text. **a, j**, Experimental and simulated bifurcation diagrams. **b-i**, highlighted P-Q spaces at specific drive levels.

# 8 Additional experimental results

To support Fig. 5 in the main text, we provide the experimental data used in panels a, b, e and f in Fig. 3. This data is obtained by sweeping up the slow voltage $V_S$ and repeating for different modulation frequencies $f_S$.

To show the observed dynamics are not only seen in the drum used in the main text, we repeated measurements on two different drums.

The second drum has a resonance frequency of 12.78 MHz, $Q$ factor of 1307 and Duffing constant of $3.82 \times 10^{17}\,\mathrm{V^{-2}s^{-2}}$. The bifurcation diagram when driving at $f_F = f_0 + 30\,\mathrm{kHz}$, $V_F = 1.5\,\mathrm{V}$ and $f_S = 20\,\mathrm{kHz}$ is shown in Fig. 11.

The third drum has a resonance frequency of 9.95 MHz, $Q$ factor of 772 and Duffing constant of $6.69 \times 10^{17}\,\mathrm{V^{-2}s^{-2}}$. This drum is driven at $f_F = f_0 + 40\,\mathrm{kHz}$, $V_F = 1.5\,\mathrm{V}$ and $f_S = 20\,\mathrm{kHz}$, which leads to the results shown in Fig. 12.

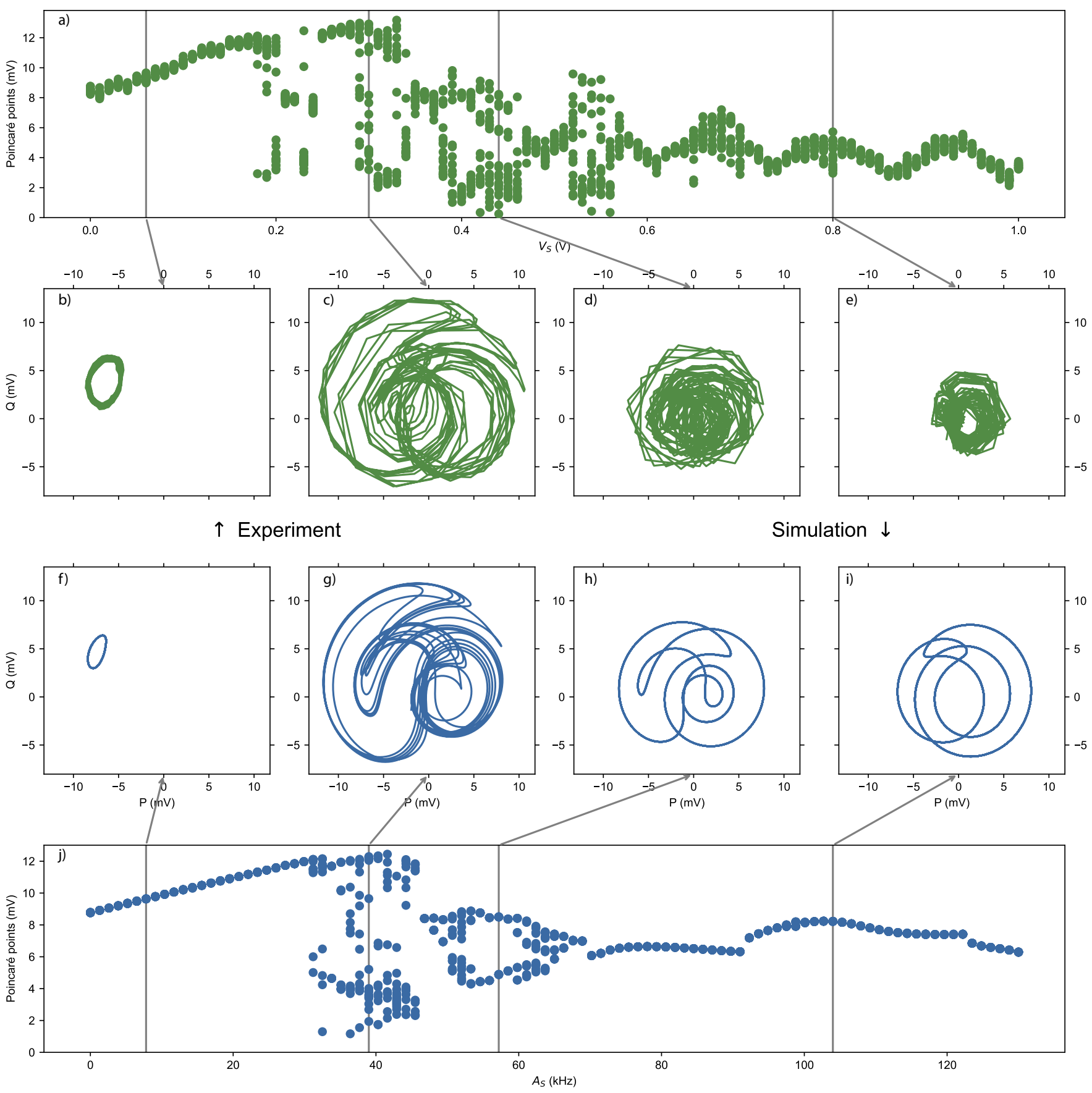


**Fig. 12** Bifurcation diagram for a second additional drum to the one in the main text. **a, j**, Experimental and simulated bifurcation diagrams. **b-i**, highlighted P-Q spaces at specific drive levels.

# References


[1] Keşkekler, A., Arjmandi-Tash, H., Steeneken, P. G. & Alijani, F. Symmetry-breaking-induced frequency combs in graphene resonators. *Nano Letters* **22**, 6048–6054 (2022).

[2] Dolleman, R. J. *et al.* Opto-thermally excited multimode parametric resonance in graphene membranes. *Scientific Reports* **8**, 9366 (2018).

[3] Rosenstein, M. T., Collins, J. J. & De Luca, C. J. A practical method for calculating largest lyapunov exponents from small data sets. *Physica D: Nonlinear Phenomena* **65**, 117–134 (1993).

[4] Melnikov, V. On the stability of a center for time-periodic perturbations. *Transaction of the Moskow Mathematical Society* **12**, 3–52 (1963).

[5] Guckenheimer, J. & Holmes, P. *Nonlinear Oscillations, Dynamical Systems, and Bifurcations of Vector Fields* Applied Mathematical Sciences (Springer, New York, 2013).

[6] Yagasaki, K., Sakata, M. & Kimura, K. Dynamics of a weakly nonlinear system subjected to combined parametric and external excitation. *Journal of Applied Mechanics* **57**, 209–217 (1990).



\end{document}